\documentclass[useAMS,usenatbib]{mn2e}
\def\kms{km\,s$^{-1}$}

\def\Ha{H$\alpha$}
\def\Hb{H$\beta$}
\def\Hg{H$\gamma$}
\def\Hd{H$\delta$}

\def\M{M$_{\odot}$}

\usepackage{graphicx}

\title[Supernova 1996al]
{The spectacular evolution of Supernova 1996al over 15 years: a low energy explosion of a stripped massive star in a highly structured environment}

\author[Benetti et al.]
{S. Benetti$^{1}$, N. N. Chugai$^2$, V. P. Utrobin$^3$, E. Cappellaro$^{1}$, F. Patat$^{4}$, A. Pastorello$^{1}$,
\newauthor 
M. Turatto$^{1}$, G. Cupani$^{5}$, R. Neuh\"auser$^{6}$, N. Caldwell$^{7}$, G. Pignata$^{8,9}$, L. Tomasella$^1$
\\
$^1$INAF - Osservatorio Astronomico di Padova, vicolo dell'Osservatorio 5, I-35122 Padova, Italy \\ 
$^2$Institute of Astronomy of Russian Academy of Sciences, Pyatnitskaya St. 48, 119017 Moscow, Russia\\
$^3$Institute of Theoretical and Experimental Physics, B. Cheremushkinskaya St. 25, 117218 Moscow, Russia\\
$^4$European Southern Observatory, Karl-Schwarzschild-Str. 2, D-85748 Garching bei M\"unchen, Germany\\
$^5$INAF - Osservatorio Astronomico di Trieste, Via G.B. Tiepolo, 11 I-34131 Trieste, Italy\\
$^6$Astrophysikalisches Institut und Universit\"ats-Sternwarte, FSU Jena, Schillerg\"a\ss chen 2-3, D-07745 Jena, Germany\\
$^{7}$Harvard-Smithsonian Center for Astrophysics, Cambridge, MA 02138, USA\\
$^{8}$Departamento de Ciencias Fisicas, Universidad Andres Bello, Avda. Republica 252, Santiago, Chile\\
$^{9}$Millennium Institute of Astrophysics, Chile\\
}

\date{Received ................; accepted ................}

\begin{document}

\maketitle

\begin{abstract}
Spectrophotometry of SN~1996al carried out throughout 15 years is presented. The early photometry suggests that SN~1996al is a Linear type-II supernova, with an absolute peak of $M_V \sim -18.2$ mag. Early spectra present broad, asymmetric Balmer emissions, with super-imposed narrow lines with P-Cygni profile, and He I features with asymmetric, broad emission components. The analysis of the line profiles shows that the H and He broad components form in the same region of the ejecta. By day $+142$, the \Ha~profile dramatically changes: the narrow P-Cygni profile disappears, and the \Ha\/ is  fitted by three emission components, that will be detected over the remaining 15 yrs of the SN monitoring campaign. Instead, the He I emissions become progressively narrower and symmetric. A sudden increase in flux of all He I lines is observed between 300 and 600 days.
Models show that the supernova luminosity is sustained by the interaction of low mass ($\sim 1.15$ \M) ejecta, expelled in a low kinetic energy ($\sim 1.6 \times 10^{50}$ erg) explosion, with highly asymmetric circumstellar medium. 
The detection of \Ha~emission in pre-explosion archive images suggests that the progenitor was most likely a massive star ($\sim25$ \M\/ ZAMS) that had lost a large fraction of its hydrogen envelope before explosion, and was hence embedded in a H-rich cocoon. The low-mass ejecta and modest kinetic energy of the explosion are explained with massive fallback of material into the compact remnant, a $7-8\,M_{\odot}$ black hole.
 \end{abstract} 
 

\section{Introduction} \label{int}
The aftermath of a supernova explosion is a bubble of gas expanding
at very high velocity. Eventually, the ejecta may impact on the
pre-existing circumstellar material (CSM), generating a shock in which
a fraction of the kinetic energy of the ejecta is converted into
radiation. The intensity of the resulting emission mostly depends on the
density of the CSM and the velocity contrast between the ejecta and
the CSM. If the density of the CSM is low, the emission
from the CSM-ejecta interaction becomes visible only after the SN luminosity has faded,
sometimes several years after the explosion. However, occasionally, the
CSM near the SN is so dense that the CSM-ejecta interaction
dominates the SN emission even at early phases.

With improved statistics and quality of observations, we are now
observing counterparts for the different scenarios, and a new taxonomy
for type II supernovae (SNII), based on the mass of the residual hydrogen envelope and the strength of the
CSM-ejecta interaction signatures, can be proposed.
The emission of classical SNII is
determined by the thermal balance in the ejecta, and CSM-ejecta
interaction is negligible (at least at early phases). Depending on
the shape of the light curve, they are sub-classified into plateau (IIP) and
linear (IIL) types. The current interpretation is that different light curves
correspond to different envelope masses, ranging from $\sim 10$ \M~in
SNIIP \citep[cf.][]{sma09} to $\sim 1$ \M~in SNIIL, due to different progenitor initial masses and mass loss histories.
When soon after the explosion the SN ejecta interact with
a dense CSM, the SN emission itself is
overwhelmed by the emission arising from the interaction \citep{chu90,che94,ter94}. Best examples of this class are
SN~1988Z \citep{tur93,are99}, with a long plateau-like light curve, and SN~1998S \citep{liu00,fas00} with a linear-like light curve. These SNe are named as type IIn.

In some SNII, the CSM-ejecta interaction contribution emerges only when the $^{56}$Co decay energy input fades. The fact that these are mostly SNIIL is consistent with our understanding that they experienced stronger mass loss during their evolutions \citep[e.g. SNe~1980K, 1979C, and 1990K;][respectively]{uom86,fes99,cap95}, hence they are expected to have higher CSM density.

Somewhat in between, a few SNe have emission line profiles which reveal the
presence of a CSM around the exploding stars from early-on, though the
ejecta-CSM interaction becomes preponderant some months later.  These are sometimes labeled SNIId, where
'd' stands for 'double' profile because of the simultaneous presence
of broad profiles from the ejecta and narrow ones from the CSM. SN~1994aj \citep[][Paper I]{ben98} and SN~1996L \citep[][Paper II]{ben99} belong to this group.

In this paper we present the study of a new SN (SN~1996al) with spectroscopic
properties very similar to those of SNIId \citep{ben96}. SN~1996al was visually discovered on July 22.71 UT by \citet{eva96}, about 30" north of
the center of its host galaxy, NGC 7689. The measured position of the SN is R.A.$=23^h$33$^m$16$^s.29\pm 0.03$; Decl.$=-54^o$04'59$".69\pm 0.05$ (J2000.0), in agreement with that reported by \citet{eva96}.\\
In view of its relative closeness and apparent brightness, an extensive observational campaign
was set-up with the ESO-Chile telescopes. Thanks to this massive effort, we have been able to follow the SN for more than 15 years, and
it has become one of the best monitored SNIIL-SNIId.\\
In this paper we will describe the physical properties of the explosion and constrain the nature of SN~1996al progenitor star.

\section{Observations} \label{obs}
Spectroscopy and imaging observations were carried out at ESO La Silla and Paranal Observatories using a
number of different telescopes and/or instruments (see Tables \ref{obs_tab} \&
\ref{spec_tab}).\\

\subsection{Optical Photometry} \label{phot}
\begin{figure}
\includegraphics[width=10cm]{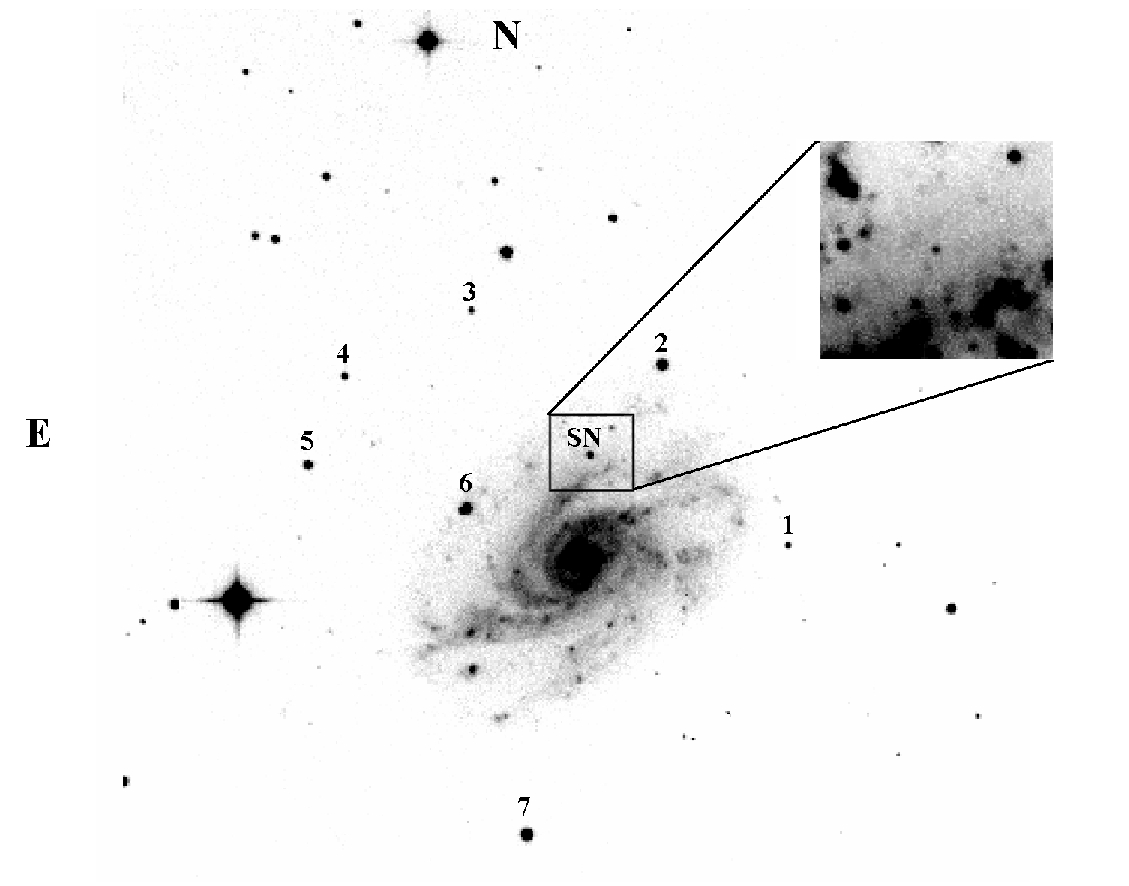}
\caption{SN~1996al in NGC~7689 and reference stars. The main image is an
$R$ frame taken on 1996 October 1 with the D1.54m+DFOSC. The
seeing is 1.4'' and the field of view is 5.9'x5.9'. The zoomed image
was obtained with VLT+FORS2 using a redshifted H$_{\alpha}$ filter on 2002
June 16. The seeing was 0.7'' and the field of view is 25''. The SN is
the source at the center of the image.} \label{sn}
\end{figure}

The CCD frames were first bias and flat-field corrected in a standard manner.
Since a fraction of data was obtained under non-photometric conditions, 
relative photometry was performed with respect to a local sequence
of field stars (see Figure \ref{sn}). Twelve photometric nights were used
to calibrate this sequence against Landolt standard stars
\citep{land}. The magnitudes and estimated errors of the local
standards are shown in Table \ref{seq}. These magnitudes were obtained by
summing the counts through an aperture, whose size varied according
to the seeing.\\

\begin{figure*}
\includegraphics[width=15cm,angle=-90]{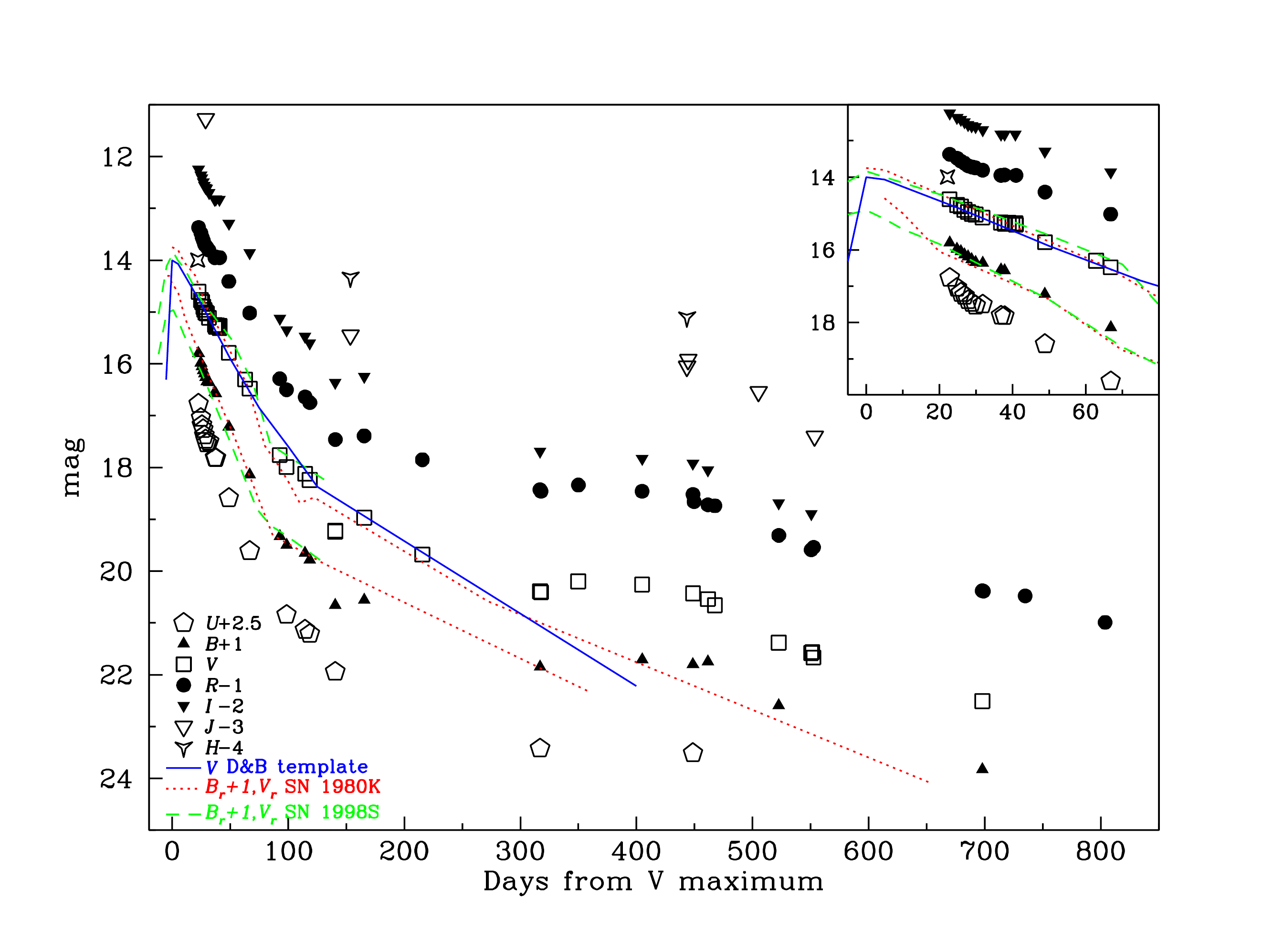}
\caption{$U,B,V,R,I,J$ and $H$ light curves of SN~1996al.  The {\it dotted}
lines represent the $B$ and $V$ light curves of SN~1980K  \citep{bar82,but82,tsv83}. The distance to the SN~1980K
host galaxy, NGC~6948, is taken from \citet{kar00}. The SN~1980K $B,V$ light curves are reported to the
distance and reddening of SN~1996al with a +0.3 mag shift, and labelled $B_r$ and $V_r$ in the figure. The {\it dashed} lines represent the $B$ and $V$ light curves of SN~1998S \citep{liu00,fas00}. The distance for the SN~1998S host galaxy, NGC~3877, is taken from \citet{wil97}. The SN~1998S $B,V$ light curves are reported to the distance and reddening of SN~1996al with a +1.5 mag shift, and labelled $B_r$ and $V_r$ in the figure
The {\it starry} symbol is the visual SN estimate at the discovery from \citet{eva96}. The best match with the template $V$ light curve of SNIIL
\citep[from][]{db85}, along with the comparisons with those of SNe~1980K and 1998S, suggest
that the maximum occurred on MJD $\sim 50265\pm 2$ days at a magnitude
$V\sim 13.9\pm0.3$.}
\label{phot_fig}
\end{figure*}

\begin{table*}
\caption{Magnitudes of the local sequence stars as identified in Figure~\ref{sn}.}
\label{seq}
\begin{tabular}{lccccc}
\hline
~~star~~ & $U$  &$B$ & $V$ & $R$ & $I$ \\
\hline   
1       & 20.36(06)&19.33(03)&18.27(02)&17.60(03)&17.06(03)\\
2$^*$    & 15.67(03)&15.84(03)&15.45(02)&15.16(02)&14.89(02)\\
3       & 21.26(11)&19.98(04)&18.52(02)&17.58(03)&16.67(03)\\
4       & 18.77(04)&18.37(05)&17.58(03)&17.09(03)&16.63(02)\\
5       & 20.37(10)&19.18(05)&17.68(03)&16.37(03)&14.89(03)\\
6$^{**}$       & 17.67(05)&17.00(05)&16.11(03)&15.53(04)&15.05(03)\\
7       & 15.70(02)&15.61(02)&15.00(02)&14.63(03)&14.31(03)\\ 
\hline
\end{tabular}

(*) For this star we have also IR magnitudes:\\
$J$=14.55(03), $H$=14.18(03), $K$=14.28(03)\\
(**) Double star: magnitudes refer to the brighter component

\end{table*}

Ideally, one would like to remove the galaxy background by subtracting
a galaxy ``template'' where the SN is absent. However, a suitable
template image was not available. Therefore, the SN magnitudes were
measured using the \textsc{IRAF}\footnote{IRAF is distributed by the National Optical Astronomy Observatories, which are operated by the Association of Universities for Research in Astronomy, Inc., under cooperative agreement with the National Science Foundation.} point spread function fitting task Daophot \citep{ste92}
and/or \textsc{ROMAFOT} package \citep{buo83} under \textsc{MIDAS}\footnote{https://www.eso.org/sci/software/esomidas/}.  These procedures allowed the simultaneous fit and subtraction of the galaxy background.  While the pixel
scales changed from one instrument to another, they were always sufficiently small to provide good sampling of the PSF (see last column, and caption of Table \ref{obs_tab}).  The supernova magnitudes with estimated internal errors are listed in Table \ref{obs_tab}, which also gives the seeing for each epoch,
averaged over the observed bands.\\
On Dec. 8, 2007 an \Ha~frame of SN~1996al was also secured with VLT+FORS2, and the frame was calibrated observing in the same night the spectrophotometric standard G158-100 \citep{oke90}. The derived \Ha~flux inside the SN~1996al PSF was found to be in excellent agreement with that derived from the subsequent FORS2 spectrum (Sect. \ref{spec}).

\subsubsection{Light curves}\label{lc}

The $UBVRIJH$ light curves are plotted in Figure \ref{phot_fig}. \\
\shortcite{pat} showed that SNII can be characterised on the basis of
the $\beta_{100}$ parameter, which quantifies the luminosity decline
rate in the first 100 days after maximum light. For SN~1996al, we
measure $\beta^{\rm B}_{100}=4.84$ \,mag\, (100d)$^{-1}$, which is
typical of the type IIL subclass, very similar for instance to SN~1980K.
However, the post maximum decline shows a short plateau-like break (see inset of Figure \ref{phot_fig}) in all
bands. This flattening is reached at day +30 in $U$ and $B$ bands, and is very short ($2-3$ days). In $VRI$ it is reached a few days later (+34d), and it lasts about one week.

The comparisons with the light curves of SNe~1980K and 1998S $V$, and the
 $V$-band template light curve of SNIIL from \shortcite{db85}, suggest that
SN~1996al reaches the $V$ maximum on MJD=$50265\pm 2$
(1996 June 30), i.e. 23 days prior to our first observation, and likely with a $V$ peak magnitude of about $13.9\pm 0.3$ ($B=13.7\pm 0.3$).
In the following, we will adopt this estimate of the $V$ maximum epoch as reference time.  The epoch of the $B$-band peak is coincident with that of the $V$ maximum. However, we stress that these estimates are subject to relatively large uncertainty, since the maxima
themselves were not observed and Type IIL SNe show a variety of 
photometric behaviours around the maximum epochs \citep{far14}. In Table \ref{lc_param}, the main parameters of the $UBVRI$ light curves
of SN~1996al are summarised.

\begin{table*}
\caption{$UBVRIJ$ light curve parameters.}\label{lc_param}
\begin{flushleft}
\begin{tabular}{lccccccc}
\hline   
\hline
 band  &post-max   &start short       &plateau  &plateau&post-plateau&  decline & decline\\
         &decline          &plateau  &duration&        &decline         &             &               \\
         &                  &from $V$max &          &             &                    &(150--300 days)&($>450$ days)\\
\hline
         &mag(100days)$^{-1}$  &days    &days  &   mag    &mag(100days)$^{-1}$&mag(100days)$^{-1}$&mag(100days)$^{-1}$\\
$U$      & 10.73            &+30      & 2    &15.01  &4.98         & $\ge 0.84$& \\
$B$      & 7.46             &+30      & 2    &15.35  &4.85         & 0.85      & 0.84\\    
$V$      & 5.55$^*$          &+34      & 7    &15.28  &4.65         & 0.89      & 0.86\\
$R$      & 5.03             &+34      & 7    &14.95  &4.38         & 0.67      & 0.67\\
$I$      & 5.12             &+34      & 7    &14.83  &4.31         & 0.95      & 0.98\\
$J$      &                  &         &      &       &             &           &1.24\\
\hline
\end{tabular}

(*) excluding the visual estimation of \citet{eva96}\\

\end{flushleft}
\end{table*}

An inflection point in the linear decline is observed at about day +150.
Between +150 and about +320 days the light curves, though not well sampled, show declines that match the $^{56}$Co decay rate (0.98 mag(100)$^{-1}$ days (Table \ref{lc_param}). From +320 to +450 days the light curves level-out to a new plateau, where the $V$ light curve slightly increases in luminosity by +0.2 mag, and $R$  by $\sim +0.1$ mag. After this plateau, the $BVRIJ$ light curves start a monotonic
decline (with some modulation in the $R$-band) with rates very similar to those observed between 150 and 300
days (Table \ref{lc_param}).

\subsubsection{Colour curves}\label{cc}

\begin{figure}
\includegraphics[width=9.5cm,angle=0]{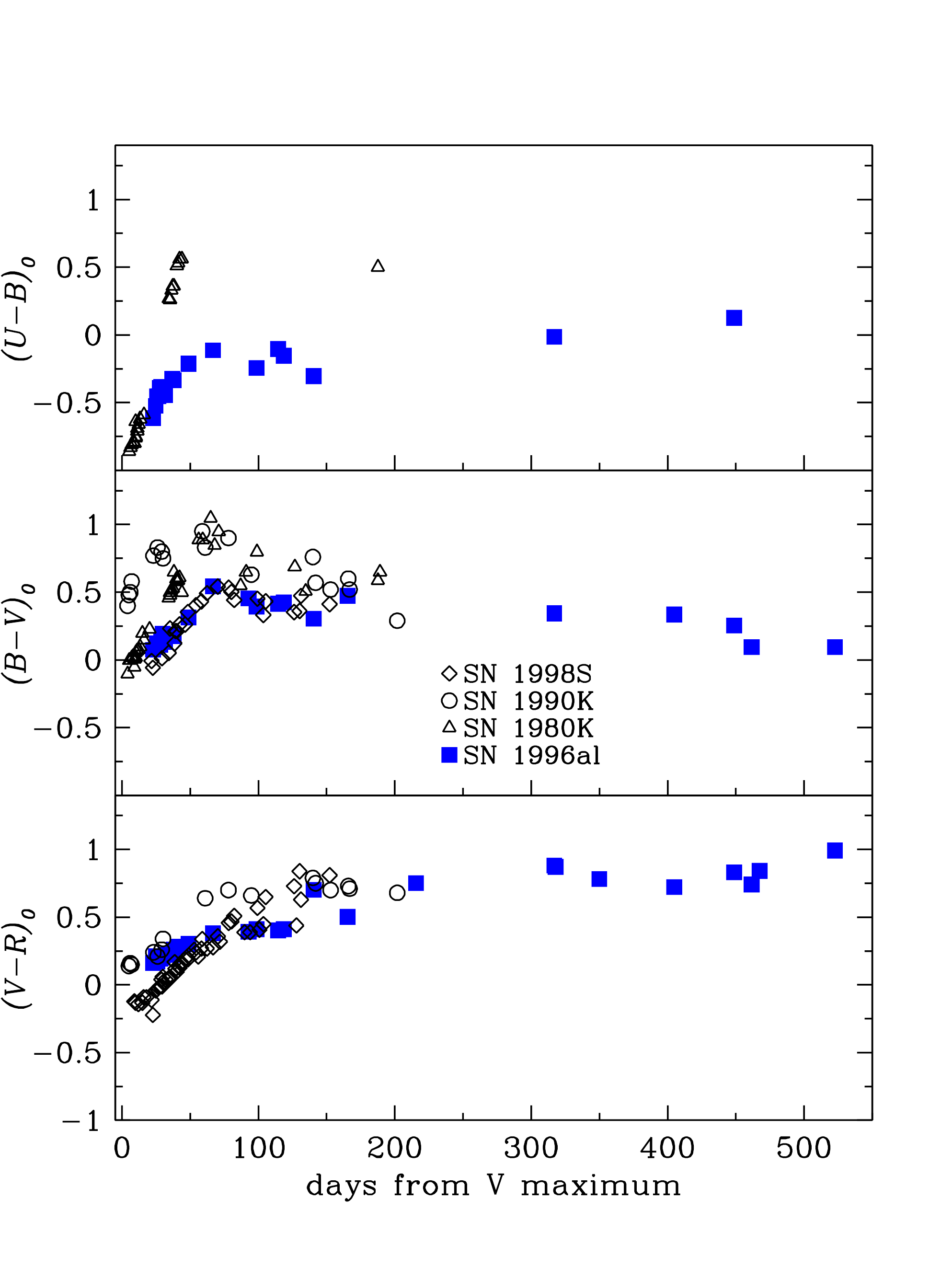}
\caption{The de-reddened colour curves of SN~1996al are compared with those of
SN~1980K and SN~1990K \citep{cap95}.
}\label{col_fig}
\end{figure}

In Figure \ref{col_fig}, the extinction corrected\footnote{using the \citet{card} extinction law, and the reddening value discussed in Sect. \ref{red}} ($U-B$)$_0$, ($B-V$)$_0$ and ($V-R$)$_0$ colour curves of SN~1996al are plotted along with those of two bright SNIIL, SNe~1980K and 1990K, and SN~1998S. \\

The early ($U-B$)$_0$ colour curve of SN~1996al is consistent with that
of SN~1980K. It shows a smooth increase reaching a value of
--0.1 mag at about 70 days. From day +140, it experiences a linear increase from --0.30 to +0.13 mag at phase 450 days, although
the colour curve sampling is poor. However, it never reaches the red
colours observed in SN~1980K (+0.5 mag).\\ 
The ($B-V$)$_0$ colour curve shows an overall similar behaviour as the
($U-B$)$_0$ colour curve. It is initially blue (+0.07 mag), then it smoothly
increases to a maximum of +0.54 mag at $\sim 70$ days, followed by a slow
decline. At late times, the ($B-V$)$_0$ colour curve
remains substantially constant at +0.34 mag with a small decrease to about
0.24 mag at about 460 days after maximum.\\

The overall appearance of the ($B-V$) colour curve is very similar to that of
SN~1998S, and not very discrepant from that of SN 1980K, although somehow displaced in phase. Also in this colour, SN~1996al never reaches the red values (($B-V$)$\sim 1$ mag) shown by SN~1980K. SN~1990K, another bright SNIIL studied by
\shortcite{cap95}, showed very red colours already at maximum. At phases close to
$\sim +160$ days, all four SNe have similar ($B-V$) colours.
Finally, the SN~1996al ($V-R$) colour curve experiences a gradual ``reddening'' with early values
similar to those seen in SN~1990K. After +100 days, the $V-R$
colour curves of the two SNe show a plateau-like feature, with a
($V-R$)$\sim 0.41$ mag for SN~1996al. Starting at about day +120, ($V-R$) further increases to $\sim 0.9$ mag, and remains thereafter almost constant.

\subsection{Reddening and distance to NGC 7689} \label{red}

The Galactic reddening towards NGC~7689 is A$_B$=0.051 mag
\citep{sch98}. The higher S/N ratio spectra of SN~1996al show narrow
absorptions due to interstellar Na\,ID, suggesting that some light extinction may occur within the parent galaxy. The mean EW of this doublet
averaged over 10 measurements is $0.60\pm 0.08$\AA. Using the
relation E$(B-V)\sim 0.16\times EW$(Na\,ID) of \shortcite{macio02} yields
E$(B-V)_{host}\sim 0.10\pm 0.05$ mag. In the following, we will adopt
for SN~1996al a total reddening of E$(B-V)_{tot}=0.11\pm 0.05$ mag.\\
From the Na\,ID interstellar lines we also derive a mean recessional velocity
of $1835\pm 68$~\kms~for NGC~7689, to be compared with the optical ($1974\pm
6$~\kms) and radio ($1970\pm 5$~\kms) velocities given by NED\footnote{http://ned.ipac.caltech.edu}. The difference (-135~\kms) is likely due to the host galaxy rotation, projected along the line-of-sight (LOS).

Adopting a value of $H_0$=73~km~s$^{-1}$~Mpc$^{-1}$ along with the
radio heliocentric radial velocity of 1967 \kms, and correcting it for Local Group
infall onto the Virgo cluster, NED gives for NGC~7689 a distance
modulus of $\mu_B=31.98$ (24.89 Mpc).

On the other hand, using the relative distance of NGC 7689 from the
Virgo cluster \citep[of 1.37,][]{kra86}, and assuming a Virgo
cluster distance of 15.3 Mpc \citep{fre01}, a distance
modulus of $\mu_B=31.61$ mag (20.96~Mpc) is derived for NGC~7689.

 The mean value of the above estimates, $\mu_B=31.80\pm 0.20$ mag (22.91 Mpc)
is adopted throughout the paper as distance modulus for NGC~7689.

\subsection{Absolute magnitudes and bolometric light curve} \label{bol_lc}

With the above assumptions for extinction and distance (using the
\shortcite{card} extinction law), we obtain $M_{\rm B} \sim -18.6\pm
0.4$ and $M_{\rm V} \sim -18.2\pm 0.3$ for the SN absolute $B$ and $V$
magnitudes at maximum. These are similar to those of the ``bright'' SNIIL
sub-class \citep[which have a $< M_{\rm B} > = -19.0 \pm 0.6$ mag, rescaled to H$_0$=73 km~s$^{-1}$~Mpc$^{-1}$;][]{pat}.

\begin{figure}
\includegraphics[width=7cm,angle=-90]{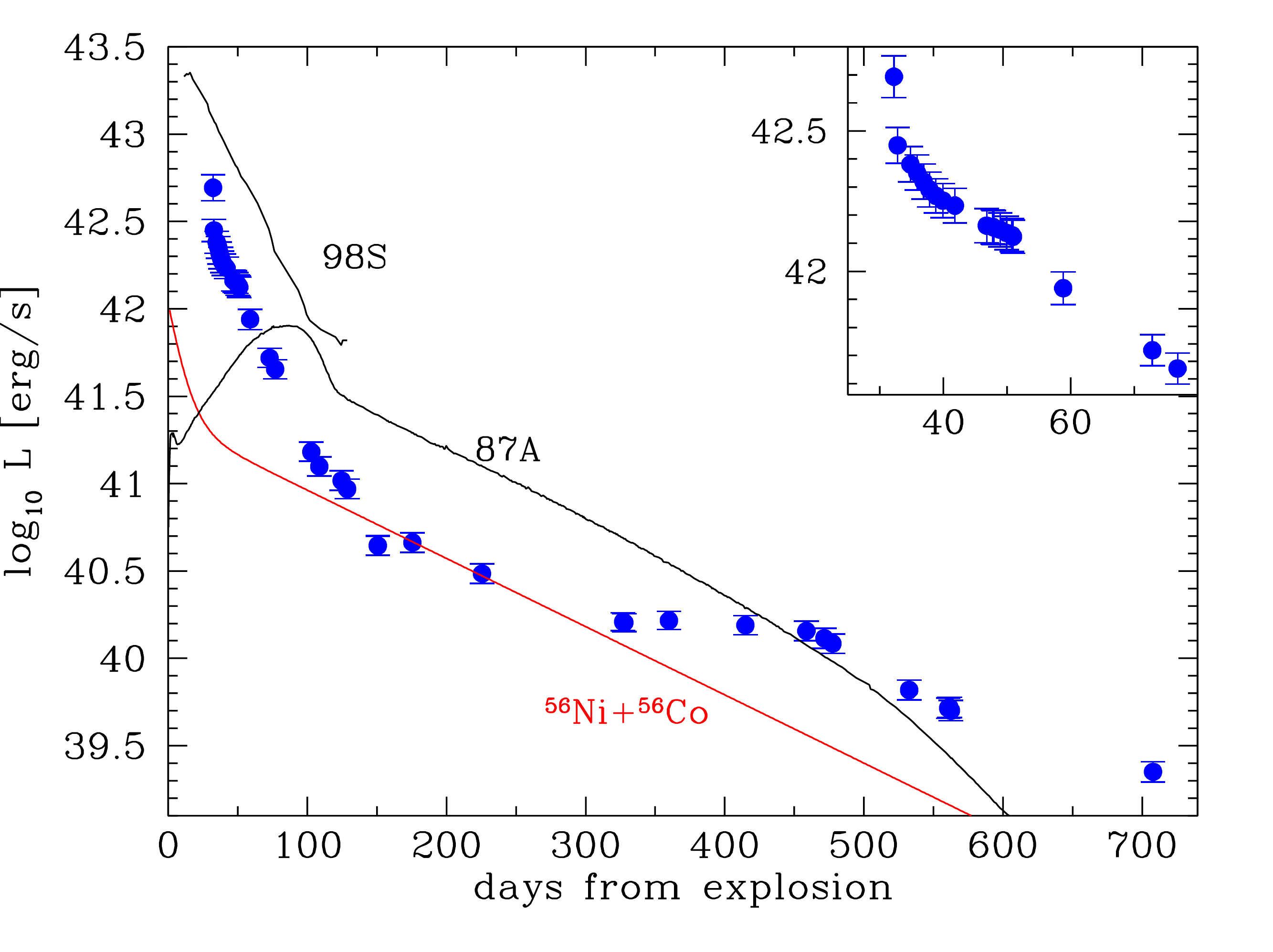}
\caption{Bolometric {\it uvoir} light curve of SN~1996al. A zoom
of the {\it uvoir} curve in phases near maximum is shown in the upper-right box. For comparison, the $^{56}$Ni+$^{56}$Co
decay model in the hypothesis of full $\gamma$-rays trapping is also reported. The rise time to maximum was assumed to be 10 days.}\label{bol_fig}
\end{figure}

In order to compute the energy budget of SN 1996al and to simplify the
comparison with theory, it is useful to determine the bolometric light
curve. Unfortunately, no observations at UV, X-ray or radio wavelengths
are available. We are then limited to derive a pseudo-bolometric
($UBVRIJHK$, or {\it uvoir}) light curve. This was obtained first adding the $UBVRIJ$ contributions and then calculating the $HK$
contributions in the phase range 153-408 days, assuming during this
period a constant $H-K$ colour. The $HK$ contribution resulted in
log$L_{UtoK}-$log$L_{UtoJ}=0.08\pm 0.01$ dex for this period. We made the assumption that the IR
contribution to the total luminosity remained constant at earlier
epochs. The bolometric curve shown in Figure \ref{bol_fig}
shares all relevant features already noted in chromatic
curves: the overall linear decay up to 150 days after explosion, including a
hint of a plateau around day +30.  It shows a break in the decay at around day +150,
then a $^{56}$Co like decay up to about day +300. This is followed by a plateau-like phase
with an almost constant bolometric luminosity (+300 to +450 days) and, finally, a new decay
with a rate close to that of $^{56}$Co.\\

\begin{table*}
\caption{Main parameter values for SN~1996al and its host galaxy.}\label{data}
\begin{tabular}{ll}
\hline
Parent galaxy           & NGC~7689     \\
Galaxy type             & SABc     $^\dag$         \\
Recession velocity  & 1970 \kms            $^\dag$\\
Distance modulus (H$_0=73$ \kms Mpc$^{-1}$)& $31.80 \pm 0.2$ mag             \\
E$(B-V)$                & $0.11\pm 0.05$ mag\\
RA$_{SN}$ (J2000.0)               &  $23^h 33^m 16^s.29\pm 0.03$         \\
Dec$_{SN}$ (J2000.0)              & $-54\degr 04'59".69\pm 0.05$           \\
Offset from galaxy nucleus     &       30''N      \\
                        &                               \\
Date of $B,V$ maxima (MJD) & $50265\pm 2.0$ (Jun. 30, 1996) \\
Magnitude at max        & $B=13.7\pm 0.3$, $V=13.9\pm 0.3$ mag,\\
$\beta^{\rm B}_{100}$   & $4.84$ mag     \\

\hline
\end{tabular}

{\dag} from NED

\end{table*}

In Sect. \ref{disc}, we will argue that most probably in SN~1996al, the CSM/ejecta interaction is the primary energy source for a large part of its evolution. However, comparing the early part of the linear tail of SN~1996al, between $\sim 150$ and 300 days after explosion, with the radioactive tail of SN~1987A, one can derive an upper limit for the $^{56}$Ni ejected in the explosion. Assuming a $^{56}$Ni mass of $0.075 \pm 0.005$ \M~for SN~1987A \citep{dan88,woo89}, and adopting for SN~1996al a scaling factor of 0.24 as suggested by Figure \ref{bol_fig}, we derive that the $^{56}$Ni ejected in the explosion of SN~1996al is $\la 0.018 \pm 0.007$ \M, where the error budget is equally dominated by the uncertainties on the SN 1996al distance, the epoch of the explosion, and the SN 1987A $^{56}$Ni mass.  We can also estimate the $^{56}$Ni mass from the observed luminosity assuming full $\gamma$-rays trapping and a rise time to maximum of 10 days. Hence, the amount of $^{56}$Co necessary to fit the tail from 100 to 300 days is $\la 0.017 \pm 0.003$ \M, fully consistent with that derived from the SN~1987A scaling.

A summary of the main parameter values inferred for SN~1996al and its host
galaxy is provided in Tab. \ref{data}.

\subsection{Spectroscopy} \label{spec}
Spectroscopic observations spanned from days +23 to +5542, with an
excellent temporal coverage during the first
two months.  Table \ref{spec_tab} lists the date (col. 1), the Modified Julian Day (col. 2), the phase
relative to $t_{V_{max}}$ (col. 3), the wavelength range (col. 4), the instrument used (col. 5), and the
resolution as measured from the FWHM of the night-sky lines (col. 6).
At some epochs, almost contemporary spectra were obtained. These were
merged to produce higher S/N spectra or wider wavelength ranges.\\

\begin{figure*}
\includegraphics[width=20cm,height=23.0cm,angle=0]{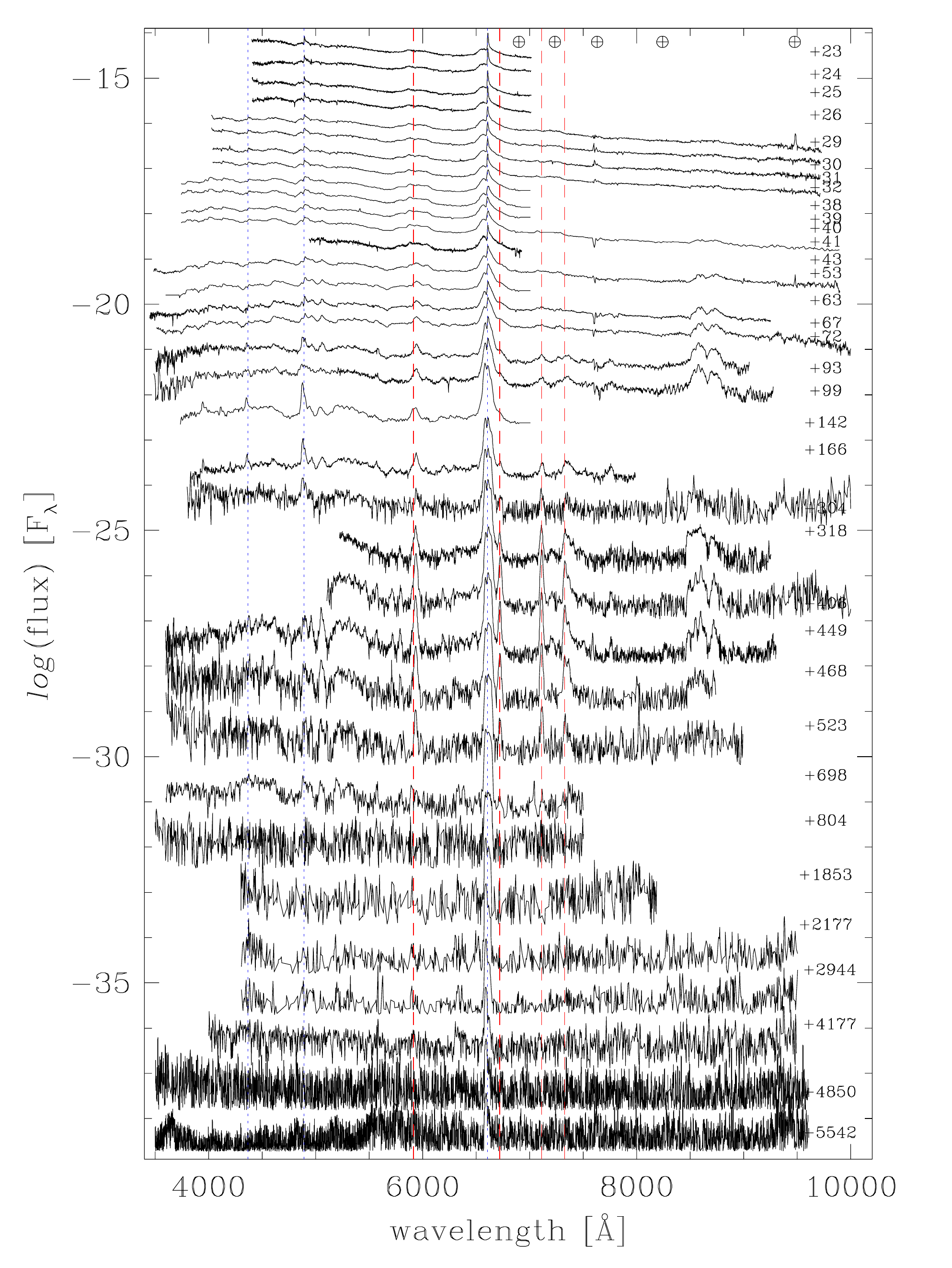}
\caption{Spectral evolution of SN~1996al. Wavelengths are in the
observer's frame. The ordinate flux scale is optimised to the first spectrum,
and all others are arbitrarily shifted downwards. In a few cases, noise spikes have been manually removed. From blue to red wavelengths, the dotted (blue) vertical lines mark the positions of \Hg, \Hb, and \Ha, respectively; while the dashed (red) lines mark the 5876 \AA, 6678 \AA, 7065 \AA~and 7281 \AA~He~I transitions. The positions of the most intense telluric absorption bands are marked.}
\label{spec_evol}
\end{figure*}

The spectra were reduced following standard MIDAS and IRAF routines. The XShooter \citep{ver11} spectra were reduced using the ESO-XShooter pipeline version 1.1.0.
One dimensional spectra were extracted weighting the signal by the variance based on the data
values and a poisson/CCD model using the gain and readout-noise
parameters. The background to either side of the SN signal was fitted
with a low-order polynomial, and then subtracted. The flux calibration and telluric absorptions modelling were achieved 
using spectra of spectrophotometric standard stars. Most spectra were
taken through a slit aligned along the parallattic angle, or normalised to spectra taken with a
wider ($\ga 5$ arcsec) slit. The flux calibration of the spectra was
checked against photometry (using the IRAF task
{\sc stsdas.hst\_calib.synphot.calphot}) and, if discrepancies occurred,
the spectral fluxes were scaled to match the photometry. On nights
with fair sky transparency, the agreement with photometry was within 15\%.  The
spectra of SN~1996al are shown in Figure \ref{spec_evol}.

\subsubsection{A brief description of the overall spectroscopic evolution}

The earliest spectra show a relatively blue continuum (T$_{bb}\sim
12000\,^o$K), once corrected for reddening and redshift. The continuum becomes
progressively redder with phase, reaching a temperature of about
$6500 - 7000\,^o$K at phases later than $\sim$60 days (see also Sect. \ref{phot_radius}). Early spectra are dominated by broad \Ha~and \Hb, and the blend of He\,I 5876\AA-Na\,ID lines.

\begin{figure}
\includegraphics[width=7cm,angle=-90]{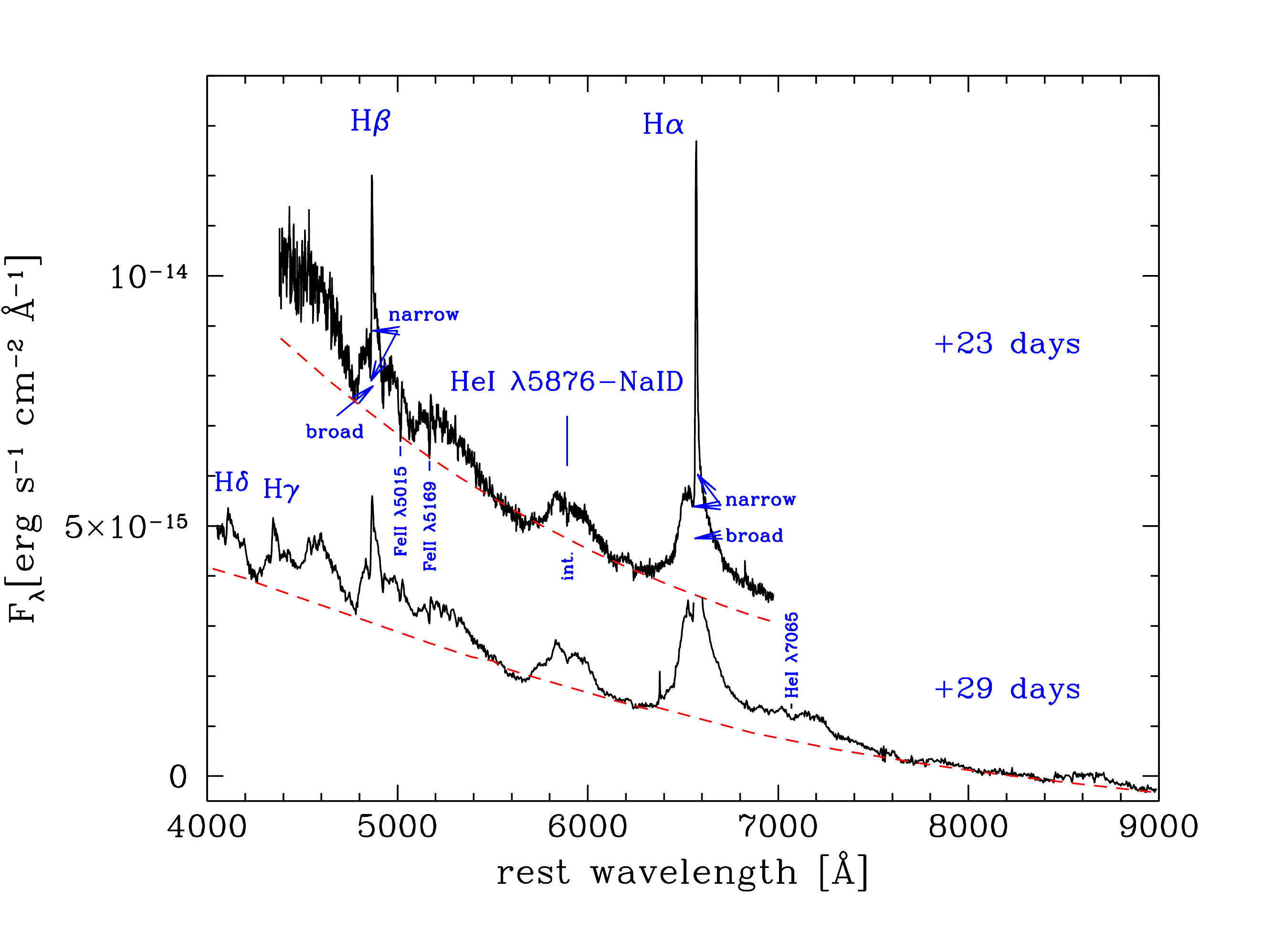}
\caption{Earliest spectra of SN~1996al corrected for redshift and reddening. The main features are identified,
together with the different components seen in \Hb~ and \Ha.
The narrow absorption (marked "int.") on the top of the He\,I\,$\lambda 5876$-Na\,ID
emission is due to Na\,ID Galactic absorption. The dashed (red) lines are blackbody fits of the spectral continuum (T=10100$^o$K and 8600$^o$K for +23 and +29 days, respectively; see Sect. \ref{phot_radius}).}\label{first}
\end{figure}

The Balmer lines show complex profiles (see Figures \ref{first} and \ref{Ha_evol}), and the decomposition of the early \Ha~profile is detailed in Figure \ref{gauss_s29}a. 
\Hb~ has a very similar profile.
The detailed evolution of the Balmer lines (in particular \Ha) is discussed in Sect. \ref{Ha_prof}.

Another intense, broad emission detected in our first spectrum is a blend of
the He\,I 5876\AA-Na\,ID lines. It has a FWHM of $\sim 260$ \AA~centered
at $\sim 5906$ \AA, though it shows a double-peaked profile (see Figure \ref{first}).
The detailed evolution of this and other unblended He lines will be discussed in Sect. \ref{He}.

A broad absorption is observed with a minimum at 5079 \AA. Assuming it is FeII 5169 \AA, it gives an
expansion velocity of $\sim 5200$ \kms.
Also, our early spectra show  narrow Fe II (5018-5169 \AA) lines
with P-Cygni profiles. These are clearly visible in the SN spectra up to phase +43 days, with a mean velocity from the absorption minima of $480\pm 150$ \kms. 

The narrow lines likely arise from a dense CSM, which was ionised either by the X-UV flash shock break-out, or by the UV photons emitted by the progenitor star (as suggested by the \Ha~emission seen around the progenitor before the explosion, cf. Sect \ref{prog}). The CSM, which is recombining at these phases, has a bulk velocity of $\sim 250$ \kms, as deduced from \Hb~narrow emission, but extends up to $\sim 2000$ \kms, as inferred from the blue edge of the narrow Balmer absorptions.

The spectrum does not change significantly during the first few weeks, with the continuum becoming progressively redder. At phase +29 days, we are able
to monitor for the first time a more extended wavelength range (see Figure \ref{first}).
Starting from the blue-end of the spectrum, we see \Hd~ and \Hg. They have complex profiles similar to those of \Hb~ and \Ha. 
Redward to \Ha, there is possible evidence for the He\,I 7065 \AA~ transition (see discussion in Sect. \ref{He}), while the He\,I 6678\AA~might be embedded in the \Ha~red wing.\\
The +41 day spectrum shows broad emissions (centered at
8531 and 8677 \AA) due to the CaII IR triplet. The FWHM of the total emission
is about 3500 \kms.
The He\,I\,5876\AA-Na\,ID feature still has a very broad  profile similar
to the emission red-ward of \Ha, which we interpret as mostly due
to He\,I 7065\AA.

The subsequent spectrum (phase +53 days) shows the appearance of other important features. The broad \Ha~profile is now better reproduced with a Gaussian profile (having a FWHM$\sim 5700$ \kms, see Table \ref{lines}), and its red wing is better fitted adding a broad (FWHM$\sim 4500$ \kms) emission centered at 6699\AA~(rest frame) in Figure \ref{gauss_s29}b. We identify this emission as the He\,I 6678\AA~transition. 

In general, the He~I lines become more symmetric and narrow with time, with their flux sharply rising after +300 day (cfr. Sect.~\ref{He}). The He~I transitions show peculiar line ratios (see Figure \ref{spec_evol} and Tables \ref{lines}, \ref{lines2}) and then disappear at about phase +700 days (see Sect. \ref{He} for a detailed discussion on the He~I lines evolution).\\
As at these early epochs the FWHM velocities of the \Ha~broad component are similar to those of the He\,I lines, it is plausible that these lines originate in the same region of the ejecta and are probably powered by the same mechanism.\\
The evolution of the \Ha/\Hb~ ratio tells that this mechanism is not photoionisation. In fact, starting from phase +43 days this ratio starts a linear increase from a value close to the case B recombination value of 2.8, to $\sim 10-12$ about 1000 days after the $V$-band maximum (see Sect. \ref{Ha_prof}).

At epochs $\ge +142$ days, the Balmer lines undergo another profound change in their profiles (see Figure \ref{gauss_s29}c). The narrow lines arising in the un-shocked CSM are not visible anymore. This could either imply that the ejecta have overtaken the bulk of the denser CSM or, more likely, that the CSM has by that time completely recombined. The \Ha~profile is now well fitted with three Lorentzians of comparable FWHM centered at 6539\AA, 6569\AA~ and 6597\AA~(see Table \ref{lines2} and Sect.~\ref{Ha_prof}). These components will be visible along the residual SN lifetime. We will refer to them as Blue, Core and Red components.  With time, the Core and Red components become weaker and weaker in comparison with the Blue one (see Figure \ref{Ha_evol}).\\
As already stressed, \Hb~becomes weaker with time with respect to \Ha~(see Sect. \ref{Ha_prof}), and the \Ha/\Hb~ratio reaches values exceeding about 15 at the latest epochs.

\section{Discussion}\label{disc}
\subsection{The progenitor star}\label{prog}

A search for archival material concerning the parent galaxy of SN~1996al returned a deep \Ha~image performed for a portion of NGC 7689 with the Rutgers Imaging Fabry-Perot instrument + CCD Camera attached at the Cassegrain focus of the CTIO 4-m Blanco telescope. The image was taken in the context of a work aimed at studying the star formation in the disk of a sample of Sa galaxies \citep{cal91}. Luckily enough, the portion of NGC 7689 imaged on September 1988 included the zone where SN~1996al appeared eight years later. Unfortunately, these were the first digital observations performed with modern CCDs and at that time CTIO was not yet equipped with a digital archive (CTIO director, private communication). For this reason, the original frames stored on a 9-track tape by the observer were lost. In order to properly place the SN location on the map we digitalised Plate 29 of Caldwell et al. and aligned it with the VLT+FORS2 \Ha~frame obtained under good seeing conditions on 2002, June 16 (see caption of Figure \ref{sn}). After the alignment, performed with about 40 sources around the SN positions, we found that a faint \Ha~region is coincident with the SN position, within an uncertainty of $\pm 0.75$". Given the relatively clean environment at that position (cfr. Figure \ref{sn}), we believe that the \Ha~emission visible in the Caldwell et al. image was indeed associated with the progenitor star of SN~1996al.\\
The \Ha~luminosity of this source measured by Caldwell et al. but not reported in their original paper, given the distance and reddening assumed in this work (see Table \ref{data}), is logL$_{H\alpha} \sim 37.28\pm 0.10$ dex. For consistency, we checked the Caldwell et al. \Ha~measurements of few H~II regions close to the SN~1996al position with those derived from the FORS2 \Ha~frame, and found a very good agreement.\\
The \Ha~pre-explosion luminosity is compared with those of some possible progenitor systems for SN 1996al in Table \ref{progenitor}. The comparison indicates that the \Ha~emission coming from the SN~1996al progenitor is similar to that expected from hot LBV stars, and lower than that measured in SN-impostor/LBVs prior to a major outburst.
This is consistent with the finding of \citet{ken80}, that H~II regions in NGC~628 with logL$_{H\alpha}< 37.90$ dex have a radius $< 200$ pc, and are supported by the radiation field of a star with a spectral type later than O4~Ia.\\
After profile decomposition (see Figure \ref{gauss_s29} and Table \ref{lines}), the narrow \Ha~component visible in the first SN spectrum has a flux of $\sim 9\times 10^{-14}$ erg s$^{-1}$ cm$^{-2}$, which provides a line luminosity of logL$\sim 39.85$ dex. This is more than two dex higher than the pre-explosion \Ha~luminosity (see Table \ref{progenitor}). The abrupt increase of the narrow \Ha~luminosity is a clear indication that the SN shock-break out or ejecta/CSM interaction ionised a much larger fraction of CSM around the progenitor star than the SN progenitor alone.
 
\subsection{The evolution of the Balmer lines}\label{Ha_prof}

The profile of Balmer lines, in particular \Ha, shows a multi-component, complex evolution which allows us to track the restless mass loss activity of the progenitor star. It also indicates an asymmetric distribution of the emitting gas. In the following, we will have a close look into the \Ha~evolution in four temporal intervals.\\

\begin{figure}
\includegraphics[width=10cm,angle=0]{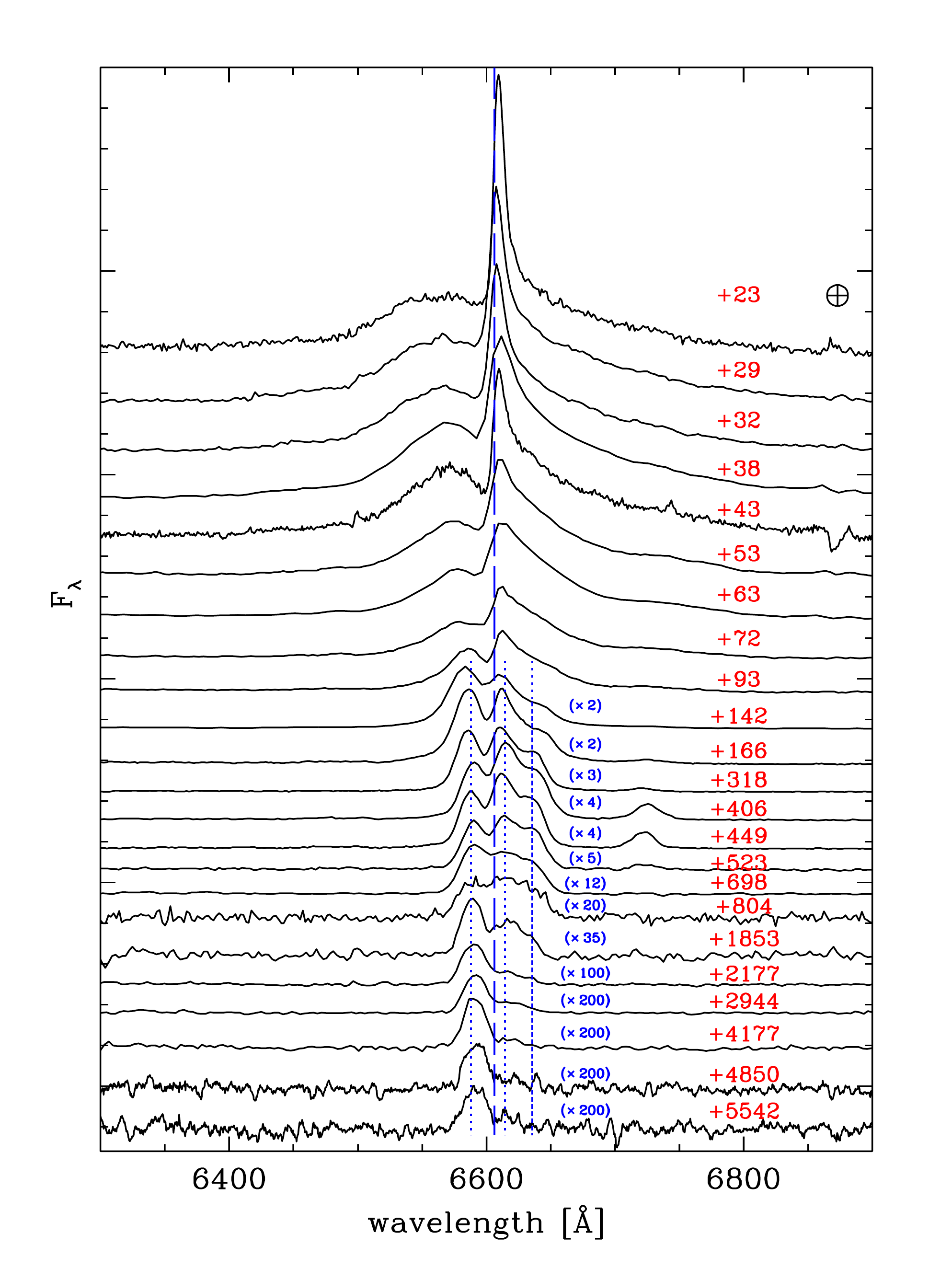}
\caption{\Ha~ profile evolution of SN~1996al. Wavelengths are in the observer's frame. The feature near 6875 \AA~in the +43 days spectrum is a residual from an poorly removed telluric band.}
\label{Ha_evol}
\end{figure}

\begin{figure}
\includegraphics[width=7cm,angle=-90]{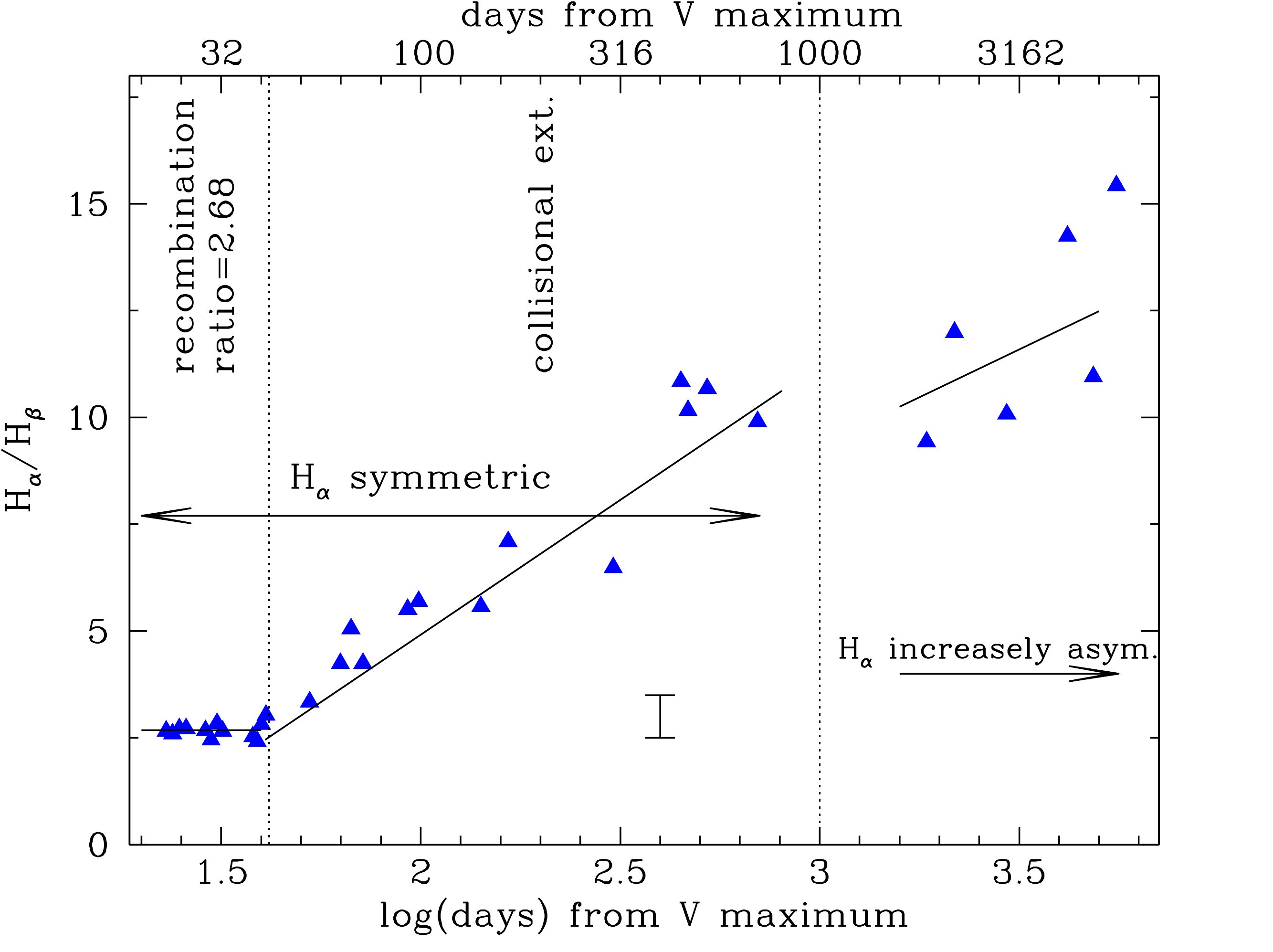}
\caption{Evolution of the \Ha/\Hb~ratio for SN~1996al. The changes of the main properties of \Ha~ are labelled. The latest points, after 1000 days, are essentially lower limits since (at these late phases) \Hb~ barely detectable. The mean error bar for the \Ha/\Hb~ratio ($\pm 0.5$) is shown.}
\label{Ha_Hb_ratio}
\end{figure}

\noindent {\bf Days +23 to +43}\\
The \Ha~profile consists of a broad Lorentzian emission with, super-posed, a narrow P-Cygni like component. The \Ha~profile does not change (see Figure \ref{Ha_evol}) until phase $\la +40$ days, which coincides with the end of the "short plateau" (see Figure \ref{phot_fig}).\\
At the same time, the expansion velocity obtained from the \Ha~FWHM of the broad component decreases from $\sim 6500$ \kms to $\sim 5600$ \kms, while the maximum velocity deduced from the \Ha~blue wing remains constant at about 12000 \kms~  (see Table \ref{lines} and Figure \ref{vel_evol}). The average wind velocity as given by the width of the narrow emission shows some scatter around the value of $\sim 440$ \kms. The mean FWHM of the narrow absorption is about 890 \kms~(see Table \ref{lines} and Figure \ref{vel_evol}).\\
The \Ha~luminosity evolution is negligible (see Figure \ref{Ha_flux}), and the \Ha/\Hb~flux ratio, which remains constant at about 2.68 (see Figure \ref{Ha_Hb_ratio}), is close to case B recombination \citep{ost89}.\\

\begin{figure}
\includegraphics[width=9.5cm,angle=0]{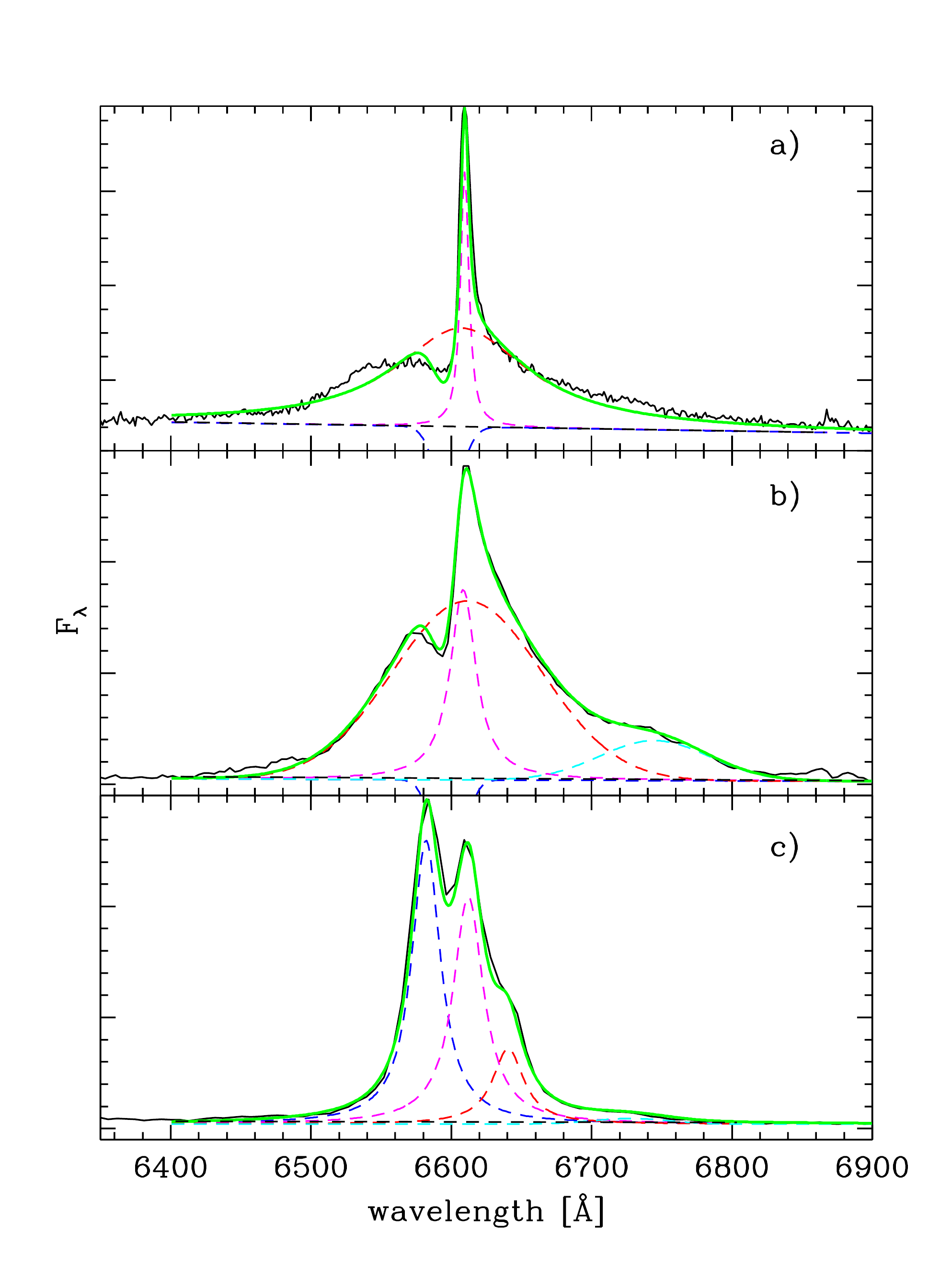}
\caption{Deconvolution of the \Ha~ profile of SN~1996al at phases +23 days (a); +53 days (b) and +142 days (c), key epochs which attest the profound modification of the line profile. Wavelengths are in the observer's frame. In panels (a) and (b) the \Ha~line profile has been fitted with a broad emission (Lorentzian in the first, Gaussian in the second; {\it dashed}, red line), a narrow Lorentzian emission component ({\it dashed}, magenta line) and a narrow Gaussian absorption component ({\it dashed}, blue line; cut down for display purpose). The feature red-ward of \Ha~in the +53 days spectrum (Gaussian; {\it dashed}, cyan line) is He\,I 6678\AA. In panel (c) the \Ha~profile has been fitted with three Lorentzians: the Blue ({\it dashed}, blue line), the Core ({\it dashed}, magenta line) and the Red component, respectively. The He\,I 6678\AA~line (Gaussian; {\it dashed}, cyan line) is also present, although now much fainter (see Sect. \ref{He}).}
\label{gauss_s29}
\end{figure}

\noindent {\bf Days +53 to +101}\\
The broad \Ha~is now best fitted by a Gaussian profile with a FWHM decreasing from  5700 \kms~to 3250 \kms, though faint, extended wings are still visible (see Figure \ref{gauss_s29}b). The maximum velocity deduced from the \Ha~blue wing remains, instead, constant at about 10500 \kms~(see Table \ref{lines} and Figure \ref{vel_evol}).\\
Again, these velocities are similar to those of the He\,I lines, hence we suppose that the lines originate in the same region of the ejecta. Since in the next Section we will demonstrate that at these phases the He\,I lines are powered by ejecta-CSM interaction, we believe that the same mechanism also supports the \Ha~flux.  At day +100, the \Ha~luminosity is about 0.3 dex lower than the \Ha~luminosity for the twin SN~1994aj at the same phase (Paper I), but is very similar to that of SN~1996L (Paper II).

The velocity of the wind, as derived from both the narrow emission and absorption features, shows a mean value of $\sim 590$ \kms.

In this period, the  \Ha/\Hb~flux ratio begins to increase, reaching a value of $\sim 6.5$ at a phase of +100 days. This could be a clue that \Ha~becomes more collisionally excited \citep{bra81}, in line with the proposed scenario (see Sect. \ref{geo}).\\

\begin{figure}
\includegraphics[width=7cm,angle=-90]{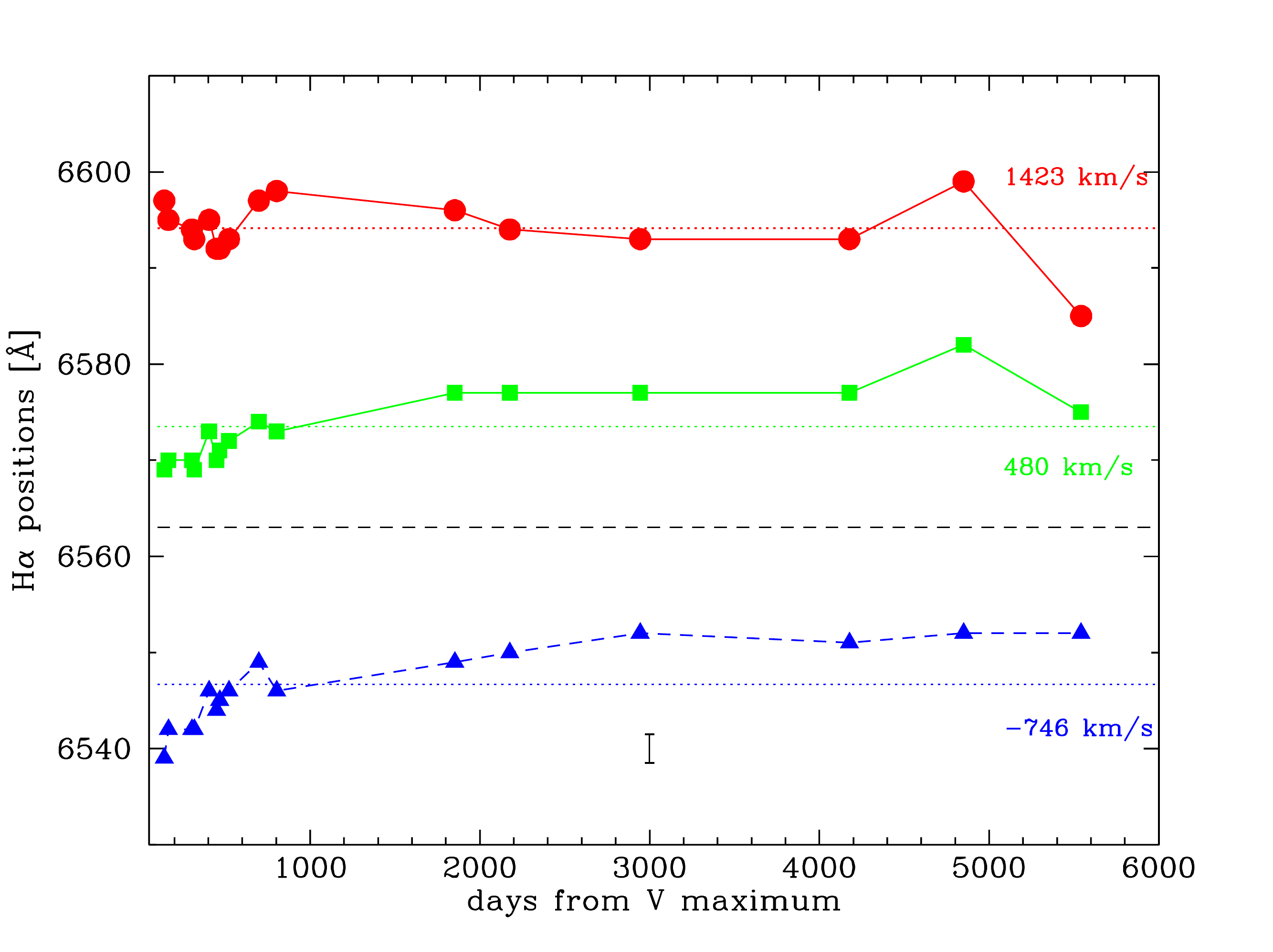}
\caption{Evolution of the positions of the three \Ha~components in the spectra of SN~1996al obtained after day $\sim 100$. The mean error bar for the component positions ($\pm 1.5$ \AA) is shown.}
\label{Ha_positions}
\end{figure}

\noindent {\bf Days +142 to $\sim$ +800}\\
From day +142, the \Ha~profile undergoes another profound change. The \Ha~profile is now better fitted by three Lorentzian components (see Figure \ref{gauss_s29}c), up to phase $+523$ days, and later on with three Gaussian components. They are centered at $\sim -1100$ \kms~(the Blue component), $270$ \kms~(the Core component) and $+1550$ \kms~(the Red component) with respect to the nominal position of \Ha.\\
The three narrower components are visible along the remaining life of the SN and their wavelengths do not change significantly with time (see Figures \ref{Ha_evol} and \ref{Ha_positions}).

Since their earliest appearance, these components show a very similar FWHM, the Red component being the broadest (FWHM$\sim 970$ \kms), followed by the Core component (FWHM$\sim 950$ \kms), while the Blue component has a FWHM$\sim 850$ \kms (see Table \ref{lines2} and Figure \ref{vel_evol}). 

Between 142 and 800 days, the flux ratios of the different components stay constant, with the Core and Blue components carrying the largest fraction of the \Ha~flux (30-40\% each).
In this lapse of time, the \Ha~luminosity shows a relatively flat and linear (in a log-log diagram) decrease up to phase $\sim +500$ days, followed by a sharper drop (by almost 1 dex) up to about +800 days.
This behaviour resembles that of the bolometric curve which, however, stops at about 700 days after explosion (see Figure \ref{bol_fig}).

We interpret these \Ha~components as the result of the interaction between a mostly spherical ejecta with an highly asymmetric CSM. The geometric configuration of the system will be discussed with more detail in Sect. \ref{geo}.\\
The \Ha/\Hb~flux ratio continues to steadily rise (see Figure \ref{Ha_Hb_ratio}) to a value of about 10.\\

\noindent {\bf Later than $+800$ days}\\
There is an observational gap between +800 and +1850 days. From day $+1853$ the \Ha~profile is again fitted by the three Gaussians mentioned before, and both the \Ha~flux and the bolometric curve show steeper evolutions.

\begin{figure}
\includegraphics[width=7cm,angle=-90]{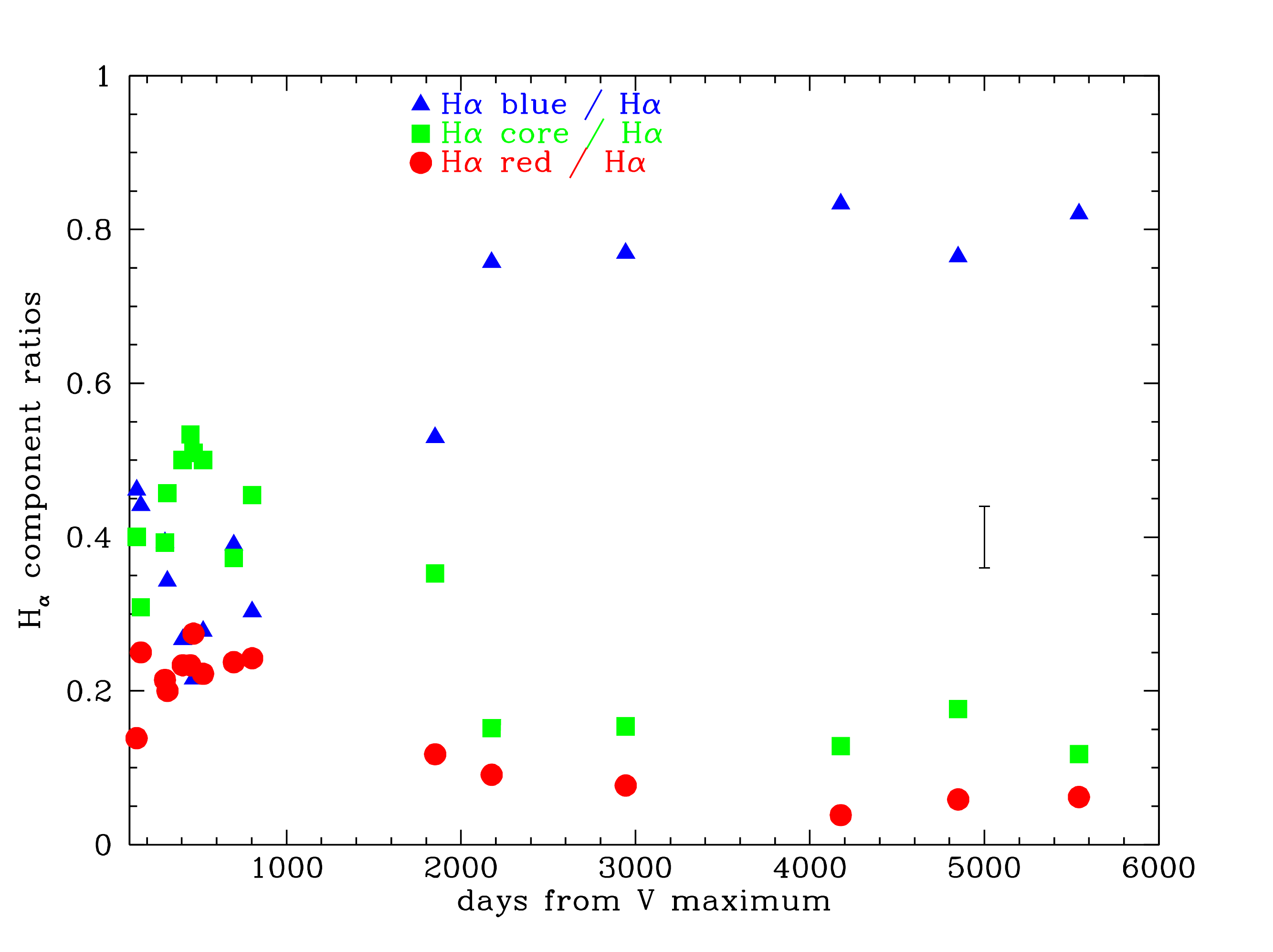}
\caption{Evolution after day $\sim 100$ of the flux ratios of the three \Ha~components in the spectra of SN~1996al. The flux of each component is divided by the total flux. The mean error bar for the flux ratios ($\pm 0.08$) is shown.}
\label{Ha_comp}
\end{figure}

While the overall flux carried out by the line dim by 0.28 dex during this time interval, the relative intensities of the three components change: the Blue component is now much stronger with respect to the other two, while the Red one remains the dimmest (see Figure \ref{Ha_comp}). At phase +2177 days, the flux ratio of the Core component decreases to the level of the Red component one, while the Blue component is carrying almost 80\% of the line flux. After this phase, the line ratios do not change very much. The sharp change in the relative flux ratios may be a signature of either a sudden dust formation in the SN ejecta, or a progressive quenching of the interaction between the ejecta and the receding part of the CSM. Most probably, as the central wavelengths of the Core and Red emission peaks do not show any evolution (see Figure \ref{Ha_positions}), both scenarios are at work. Also the Blue component does not show any shift from the central position. For this reason, if dust is forming, it condensates in the inner part of the ejecta or in the clumps where the Core and Red \Ha~components originate.

Starting from $\sim 2900$ days after the explosion, the \Ha~luminosity (see Figure \ref{Ha_flux}) decreases below the luminosity emitted by the precursor star. This, along with its distorted profile, proves that most of the \Ha~flux comes from the SN itself and that the ambient contamination is by now negligible. Moreover, this is a clue that the \Ha~source detected before the explosion was indeed associated to the precursor star and that it has been swept up by the SN ejecta.

\subsection{The He\,I lines}\label{He}

\begin{figure}
\includegraphics[width=9cm,angle=0]{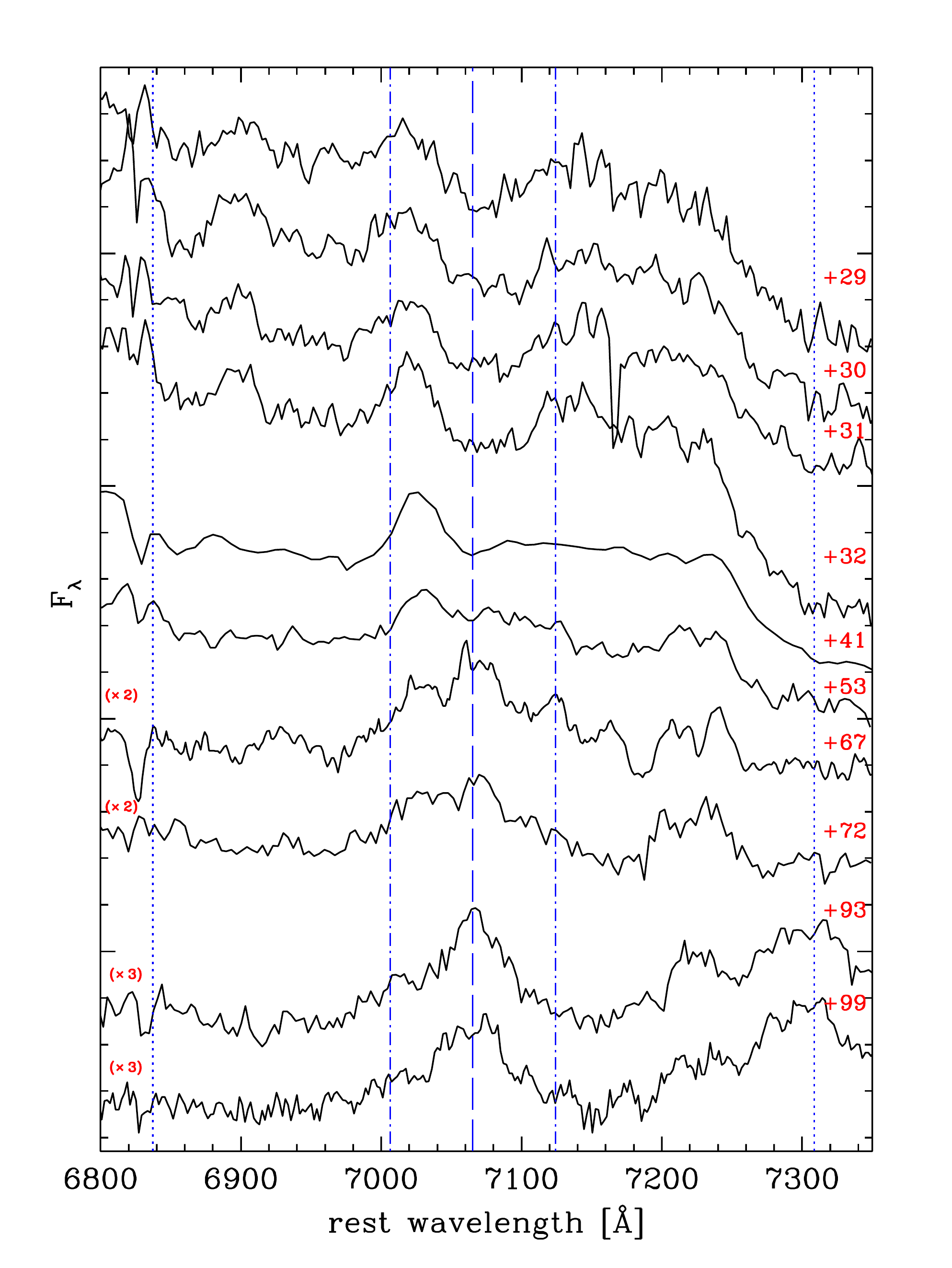}
\caption{Evolution of the He\,I~7065 \AA~line profile during the first 100 days. The {\it dashed} line marks the He\,I-7065\AA~rest frame position, while the {\it dotted} and {\it dash-dotted} lines show the expansion velocities of $\pm 10000$ \kms~and $\pm 2500$ \kms, respectively.}
\label{hei7065a}
\end{figure}

\begin{figure}
\includegraphics[width=9cm,angle=0]{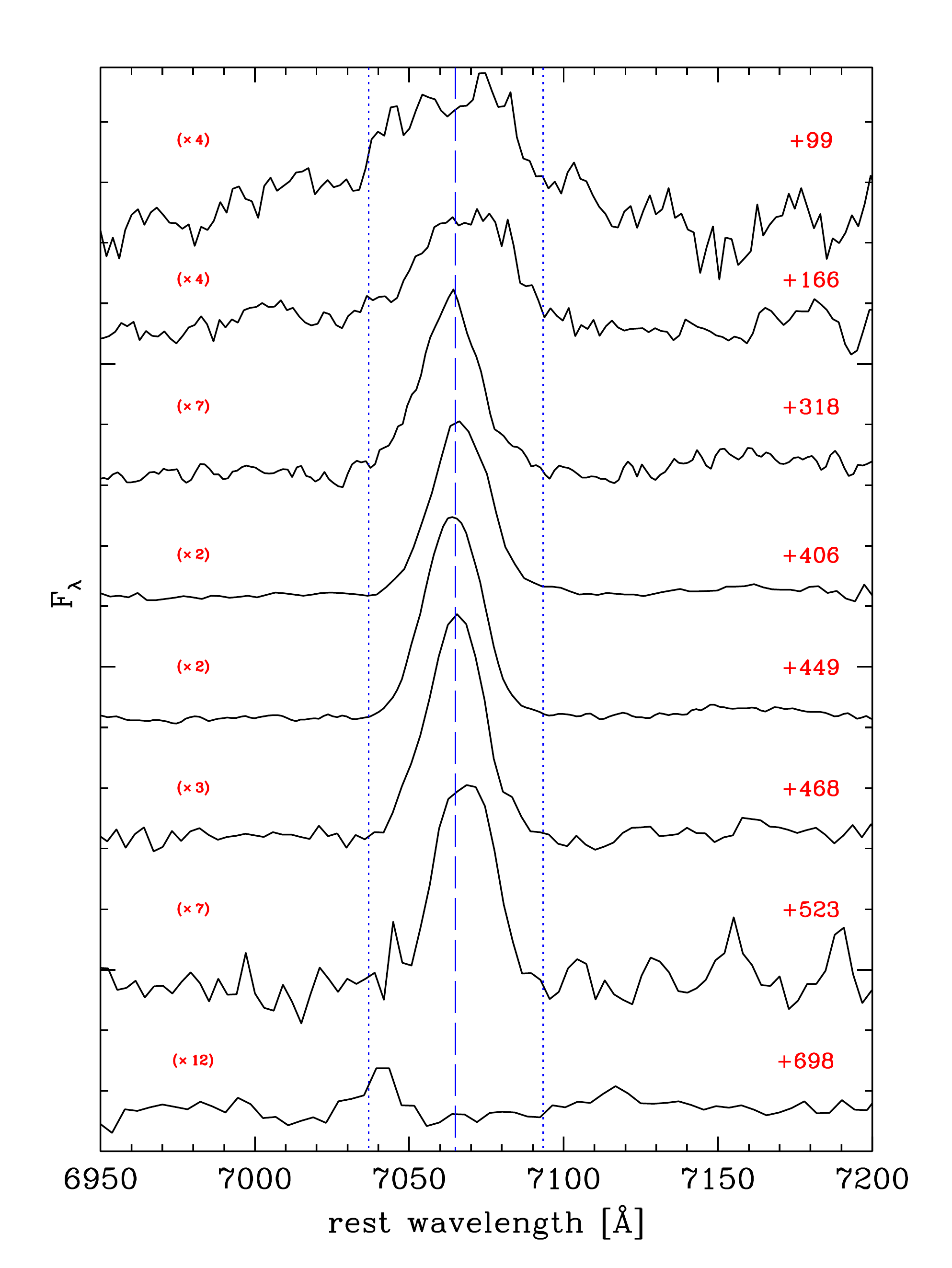}
\caption{Evolution of the He\,I~7065 \AA~line profile after 100 days. The dashed line marks the He\,I-7065\AA~rest frame position, while the dotted lines indicate the expansion velocity of $\pm 1200$ \kms.}
\label{hei7065b}
\end{figure}

In our first spectra, the best evidence for the presence of helium is the broad, asymmetric bump (FWHM$\sim 11000$ \kms) centered at about 7065 \AA~(rest frame) that can be identified as He\,I 7065 \AA~(see Figure \ref{first}). Figures \ref{hei7065a} and \ref{hei7065b} show the complete evolution of He\,I 7065\AA, the only helium transition which is not contaminated by close-by strong lines. In fact, this is not the case for He\,I~5876 \AA, which could be heavily contaminated by the nearby Na\,ID doublet. An estimate of the relative contribution of two lines throughout the complete spectroscopic evolution of SN 1996al is a very difficult task.
The last phase in which the He\,I~lines are unequivocally detected is +523 day (see Figure \ref{spec_evol}).\\

\begin{figure}
\includegraphics[width=9cm,angle=0]{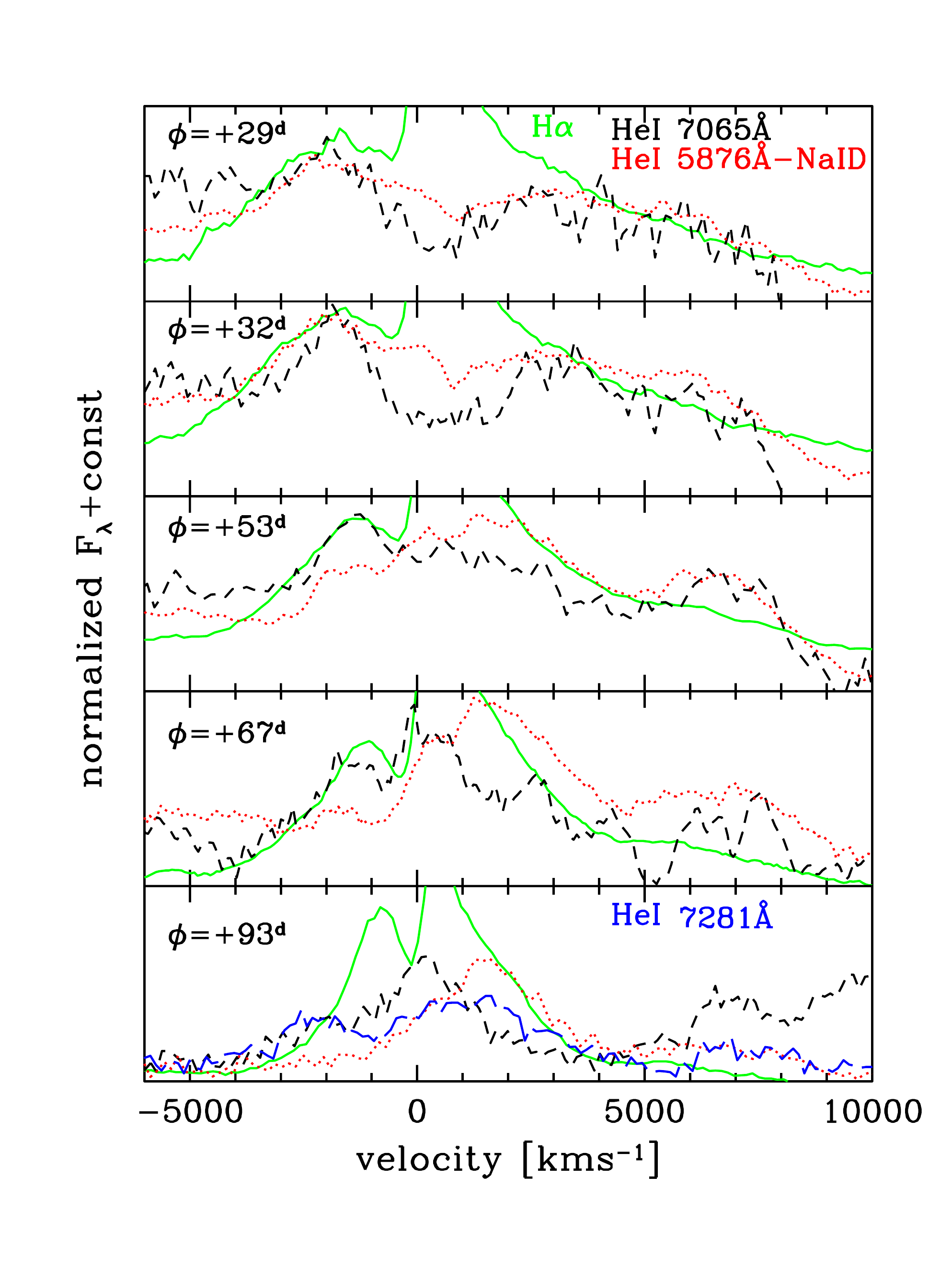}
\caption{Comparison in the velocity space of the profile evolution of the He\,I~5876 \AA~- Na\,ID feature (reference wavelength is 5876 \AA, dotted, red lines) with those of the He\,I 7065 \AA~feature (short-dashed, black lines) and \Ha~solid, green lines. In the last phase (+93 days), we also show the profile of He\,I 7281 \AA~feature (long-dashed, blue line).}
\label{He_comp}
\end{figure}

In Figure \ref{He_comp}, we compare the profiles of He\,I~5876 \AA~- Na\,ID  with the He\,I 7065 \AA~ line, during the first 100 days of the SN evolution. In the first two spectra (+29 and +32 days) they are very similar, both in the overall profile and velocity limits. The lines show a double peaked profile with the blue one peaking at $\sim -2000$ \kms and the red one at $\sim 3500$ \kms, with the red wing extending far to higher velocities ($\sim 9000$ \kms) than the blue one. Thanks to the similarity of the profiles, we may safely conclude that at phases +29, and +32 days the He\,I~5876 \AA~- Na\,ID feature is almost entirely due to He. Figure \ref{He_comp} shows that, at least during the first month, the broad wings of the \Ha~profiles are similar to those of He\,I lines, suggesting that the three lines form in the same, asymmetric region.

\begin{figure}
\includegraphics[width=7cm,angle=-90]{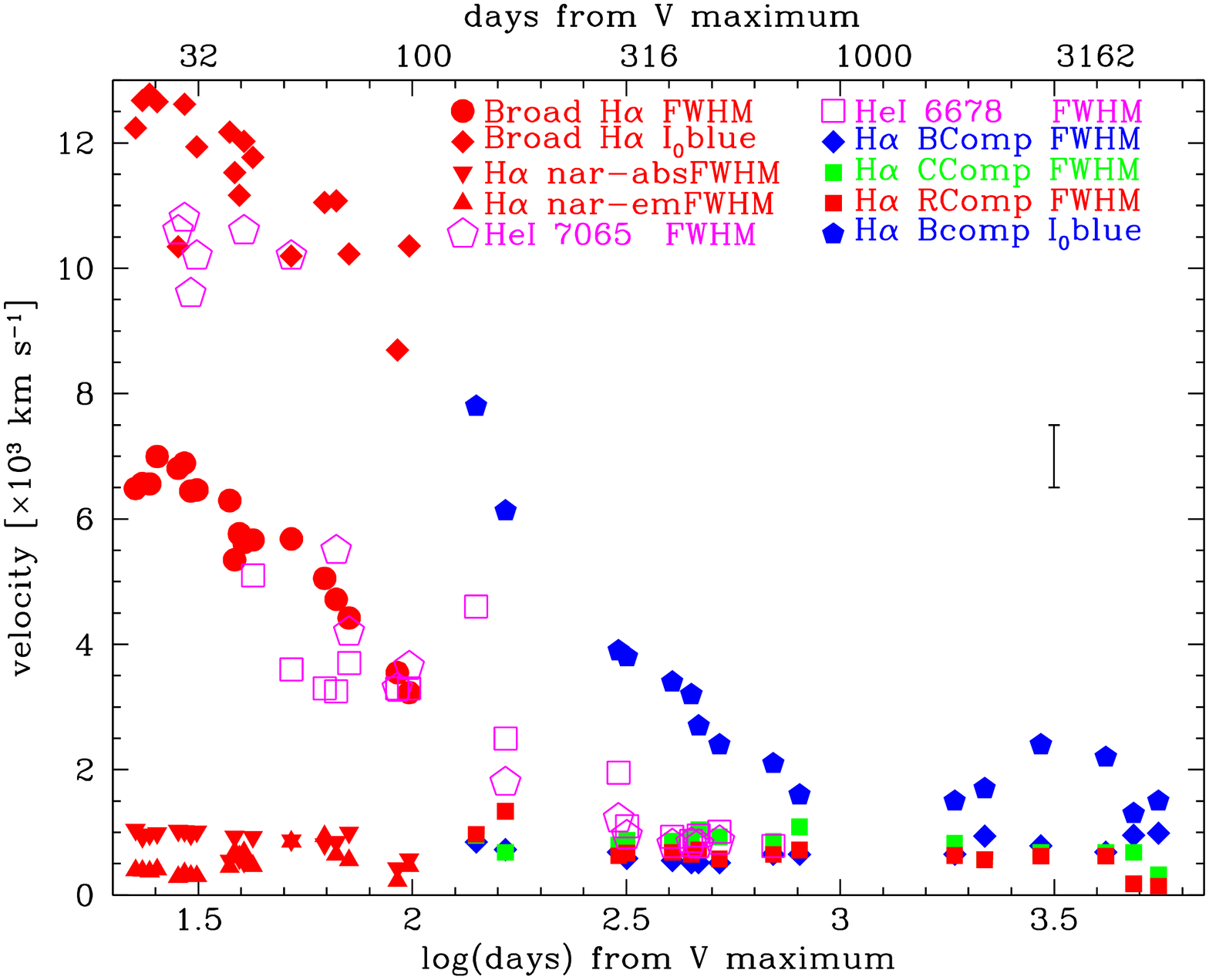}
\caption{Velocity evolution of the \Ha~components and the He\,I 6678-7065\AA~ transitions. The mean error bar ($\pm 500$ \kms) for the broader lines is shown. For the narrowest features (those with FWHM$\la 2000$ \kms), the error bar is smaller than the symbol.}
\label{vel_evol}
\end{figure}

\begin{figure}
\includegraphics[width=7cm,angle=-90]{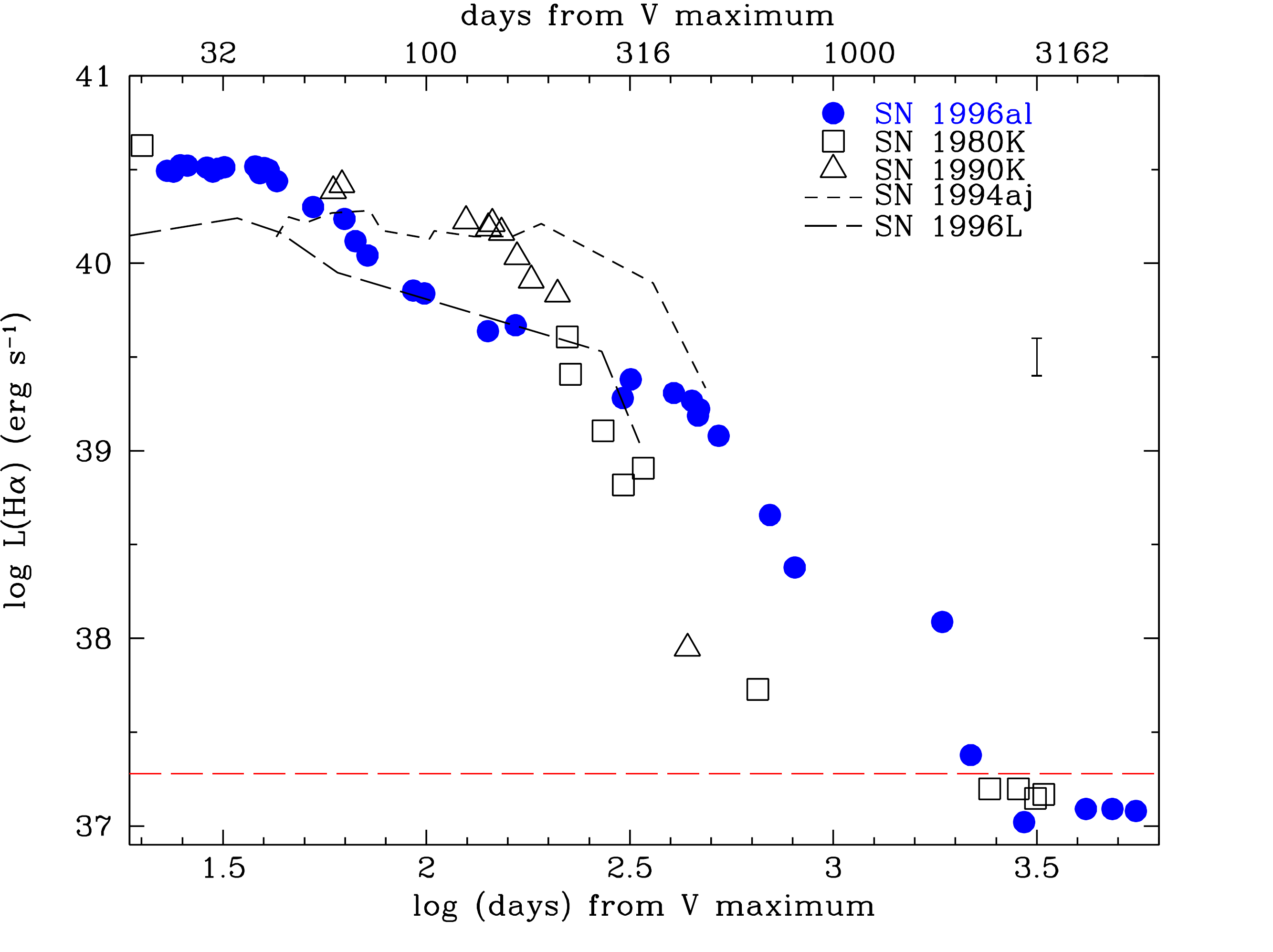}
\caption{\Ha~ luminosity evolution of SN~1996al compared with those of SNe 1980K \citep{uom86}, 1990K \citep{cap95}, and the twin SNe 1994aj (Paper I) and 1996L (Paper II). The mean error for the log~(L\Ha) of $\pm 0.10$ dex is reported. The horizontal {\it long-dashed} (red) line marks the \Ha~flux emitted by the precursor star. The \Ha~luminosities here reported are the total \Ha~flux measurements, which may be slightly different from the sum of the fitting components reported in Tables \ref{lines} and \ref{lines2} and resulting from the deconvolution of the entire line profile. The \Ha~luminosity on day $+4178$ is the average between the flux measured from the FORS2 spectrum and that derived from the calibrated \Ha~image.}
\label{Ha_flux}
\end{figure}

Figure \ref{hei7065a} also shows that the profile of the He I 7065 \AA~emission remains unchanged up to +53 days after the $V$ maximum, with the blue bump dominating the line profile. By this time, the profile of He\,I~5876 \AA~- Na\,ID feature starts to deviate from that of He I 7065 \AA~(see Figure \ref{He_comp}) with a broad bump emerging slightly redder than the Na\,ID rest wavelength position. Apart from this difference, the two He I lines and the broad \Ha~still share the same basic characteristics (see third panel from top in Figure \ref{He_comp}).

\begin{table}
\caption{Main parameter values for SN~1996al progenitor systems}\label{progenitor}
\begin{tabular}{llll}
\hline
Name     & Spectral type    & logL$_{H\alpha}$ & Reference    \\
\hline
SN 2009ip  &SN-impostor/LBV& 38.71$^{\dag}$ & 1 \\
SN 2000ch &SN-impostor/LBV& 38.48$^{\dag\dag}$& 2\\
UGC2773-2009OT&SN-impostor/LBV&38.38*& 3\\
$\eta$ Car & LVB/O5.5III-O7I& 37.55** & 4\\
AG Car &LBV/B2/3Ib-WN11& 37.54{\ddag}&5\\
S Dor & LBV/A5Iaeq C & 37.04${\ddag\ddag}$& 4 \\
HR Car & LBV/B7Ve C & 36.41${\star}$ & 4\\
$\zeta^1$ Sco& LBV/B1Ia-Oek C & 35.32${\star\star}$& 4 \\
SN~1996al prog& --- & 37.28& This paper \\
\hline
\end{tabular}
\begin{flushleft}
{\dag} Average values of three epochs (2010/10/06; 2011/09/02; 2011/09/24);
assumed $\mu = 31.55$ mag and E($B-V$)=0.019 mag from \citet{smi10}.\\
{\dag\dag} Average of two epochs (2000/05/06; 2009/05/19);
assumed $\mu = 30.17$ mag and E($B-V$)=0.013 mag from \citet{pas10}\\
{*} Average of three epochs (2012/07/20; 2009/09/05; 2010/01/26);
assumed $\mu = 28.95$ mag and E($B-V$)=0.5 mag (only Galactic) from NED (http://nedwww.ipac.caltech.edu)\\
{**}Measurements from the 2009/04/16 epoch; assumed distance of 2.3 kpc from \citet{wal12} and E($B-V$)=0.55 mag from \citet{hum12}; spectral type from SIMBAD (http://simbad.u-strasbg.fr/simbad/)\\
{\ddag} Average over eight years of measurements; assumed distance of 6 kpc from \citet{hum89,hoe92} and E($B-V$)=0.65 mag from \citet{sho96}; spectral type from SIMBAD\\
${\ddag\ddag}$ Average of four epochs (2008/02/04; 2009/02/08; 2009/09/15; 2010/08/03); assumed distance go 49.97 kpc from \citet{pie13} for LMC and E($B-V$)= 0.12 mag from \citet{mas00}; spectral type from SIMBAD\\
${\star}$ Average of two epochs (2008/12/13 and 2010/04/01); assumed distance of 5 kpc and E($B-V$)=0.9 mag from \citet{van91}; spectral type from SIMBAD\\
${\star\star}$ Average of two epochs (2009/10/19; 2010/07/04); assumed distance of 2 kpc and E($B-V$)=0.66 mag from \citet{cro06}; spectral type from SIMBAD\\
 ---\\
1= \citet{pas13}

2= \citet{pas10}

3= Pastorello et al., in preparation

4= \citet{ric12}; Richardson, private communication

5= \citet{sta01}

6= \cite{her08}

\end{flushleft}
\end{table}

Starting from phase +67 days, the He I 7065 \AA~ line shows a new component centered at the rest position, with the blue and red bumps still being visible. 
Also the He\,I~5876 \AA~- Na\,ID feature has an overall similar profile, with an increasingly contribution from the Na\,ID.\\
Afterwards, (on day +93 and +99) the He I 7065 and 7281 \AA~features become progressively more symmetric, are centered at the rest position and reach terminal velocities of about 2500-3000 \kms~(see bottom spectra in Figure \ref{hei7065a}). The He\,I~5876 \AA~- Na\,ID profile is different, indicating an increased contribution by Na\,ID  (see below). 
On the contrary, the terminal velocities of the \Ha~broad component are higher, implying different zones of formation. 
The broad \Ha~component is still coming from the outer, fast-moving region of the ejecta whose emissivity strongly depends on CSM interaction, while the He-Na\,ID lines mostly form in the slow expanding, spherically symmetric inner ejecta.\\
In order to explain the  broad He I line profiles at early times, we may conceive two possible alternatives: either some $^{56}$Ni was ejected at high velocity ($\sim 10000$ \kms), or the He\,I lines were supported by the interaction of the fast expanding symmetric ejecta with an asymmetric CSM. Since starting from about 2 months past explosion the He\,I lines become symmetric (the inner geometry of the innermost layers appears to be symmetric), we believe that the interaction scenario is the most likely.

\begin{figure}
\includegraphics[width=8.5cm,angle=0]{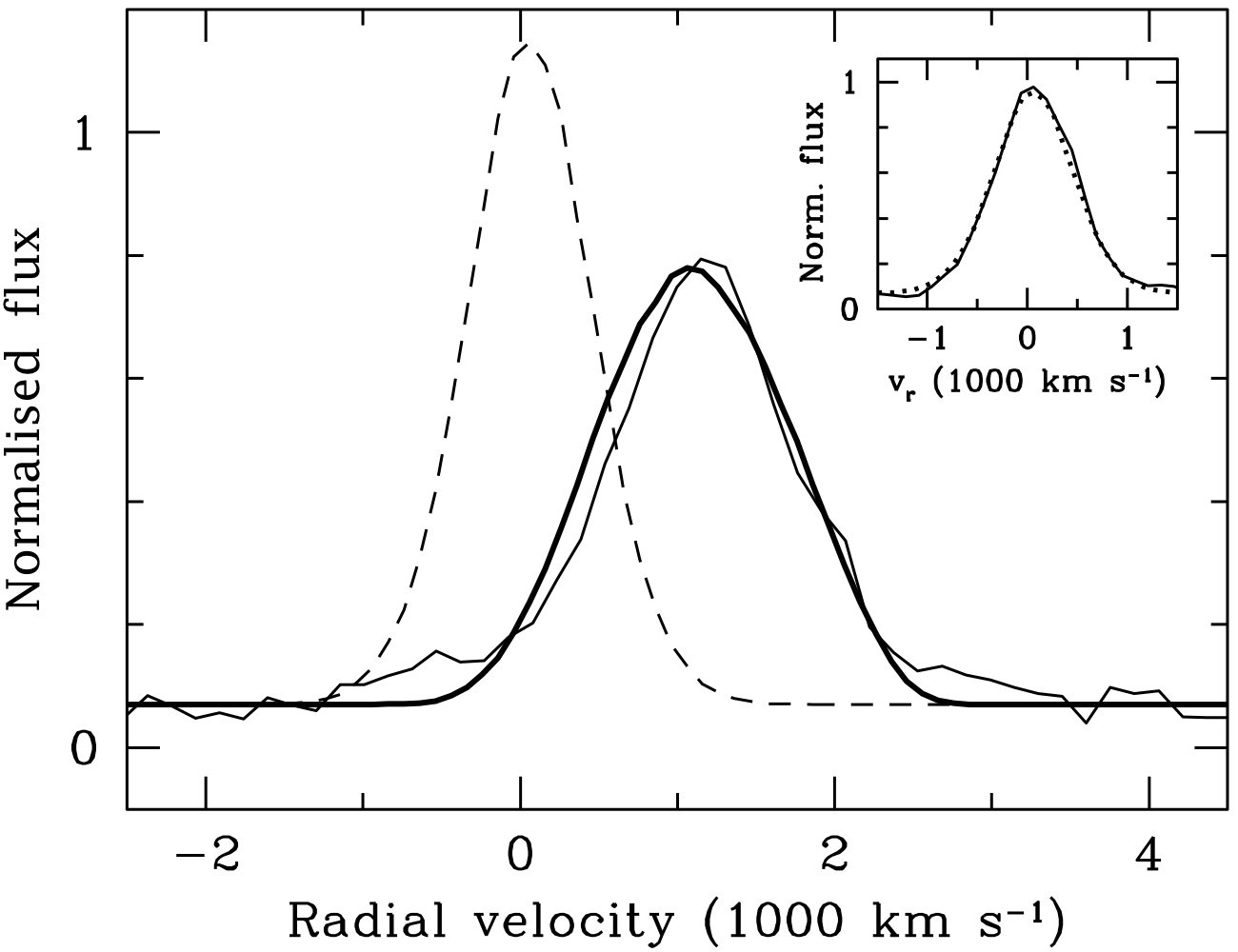}
\caption{
Effect of the non-local scattering of the He\,I 5876\,\AA\ emission by the Na\,ID in SN~1996al.
The observed spectrum on day +406 ({\it thin solid} line) is compared with
   the model spectrum ({\it thick solid} line).
The latter is the result of the scattering of the mock-up He\,I 5876\,\AA\
   emission ({\it dashed} line) in the optically thick Na\,ID lines.
The scaled mock-up He\,I 5876\,\AA\ profile ({\it dotted} line) is compared with
   the observed He\,I 7065\,\AA\ line in the {\it inset}.
The radial velocity scale refers to the He\,I 5876\,\AA\ line.
}
\label{fig:fd123}
\end{figure}

The disappearance of the He\,I 5876\,\AA\ emission line after about day +100  is {\it prima facie} at odds with the presence of the pronounced He\,I 7065\,\AA\ emission.
The clue can be found in the coeval emergence of the Na\,ID emission. 
In fact, in a co-moving frame the He\,I 5876\,\AA\ photon emitted by the SN ejecta 
can be redshifted and scattered into the Na\,ID.
As a result we see the Na\,ID feature instead of the He\,I 5876\,\AA\ emission.
The mechanism can be explored through a model based on a Monte-Carlo simulation  (Figure~\ref{fig:fd123}). 
The underlying model assumes freely expanding spherical ejecta with
   the He\,I 5876\,\AA\ emissivity distribution recovered from the
   He\,I 7065\,\AA\ profile (Figure~\ref{fig:fd123}, inset).
The adopted He\,I 5876\,\AA\ flux is 1.2 times larger than that of He\,I 7065\,\AA, corresponding to the case of collisional excitation 
   dominating over recombination \citep{alm89}. 
The optical depth of the Na\,ID is assumed to linearly fall with increasing
   velocity from $\tau(5890) = 20$
   at the SN center ($v=0$), to zero at $v = 1800$\,km\,s$^{-1}$. 
The relative fractions of the Rayleigh and spherical scattering are determined
   by the Hamilton parameter $E_1$ which, in the representation of
   \citet{cha60}, is 1/2 for Na\,I 5890\,\AA\ and zero for Na\,I 5896\,\AA.
In the optically thin limit, therefore, the scattering phase function for
   Na\,I 5896\,\AA\ line is spherical, while for Na\,I 5890\,\AA\ line the
   phase function is the superposition of the spherical (1/3) and Rayleigh
   scattering (2/3).
The latter is taken into account for optical depth $\tau(5890) \leq 1$.
For $\tau(5890) > 1$, the scattering is assumed to be spherical. This
   approximately accounts for the scattering spherization during the
   multiple local scattering.
It is reassuring that the simulated spectrum (Figure~\ref{fig:fd123}) from this simple model provides an excellent  fit to the observed emission line.  

The evidence of the non-local scattering of He\,I 5876\,\AA\ photons into Na\,ID implies that the late time He\,I line originates in the inner layers of the SN ejecta. 

\begin{figure}
\includegraphics[width=9cm,angle=0]{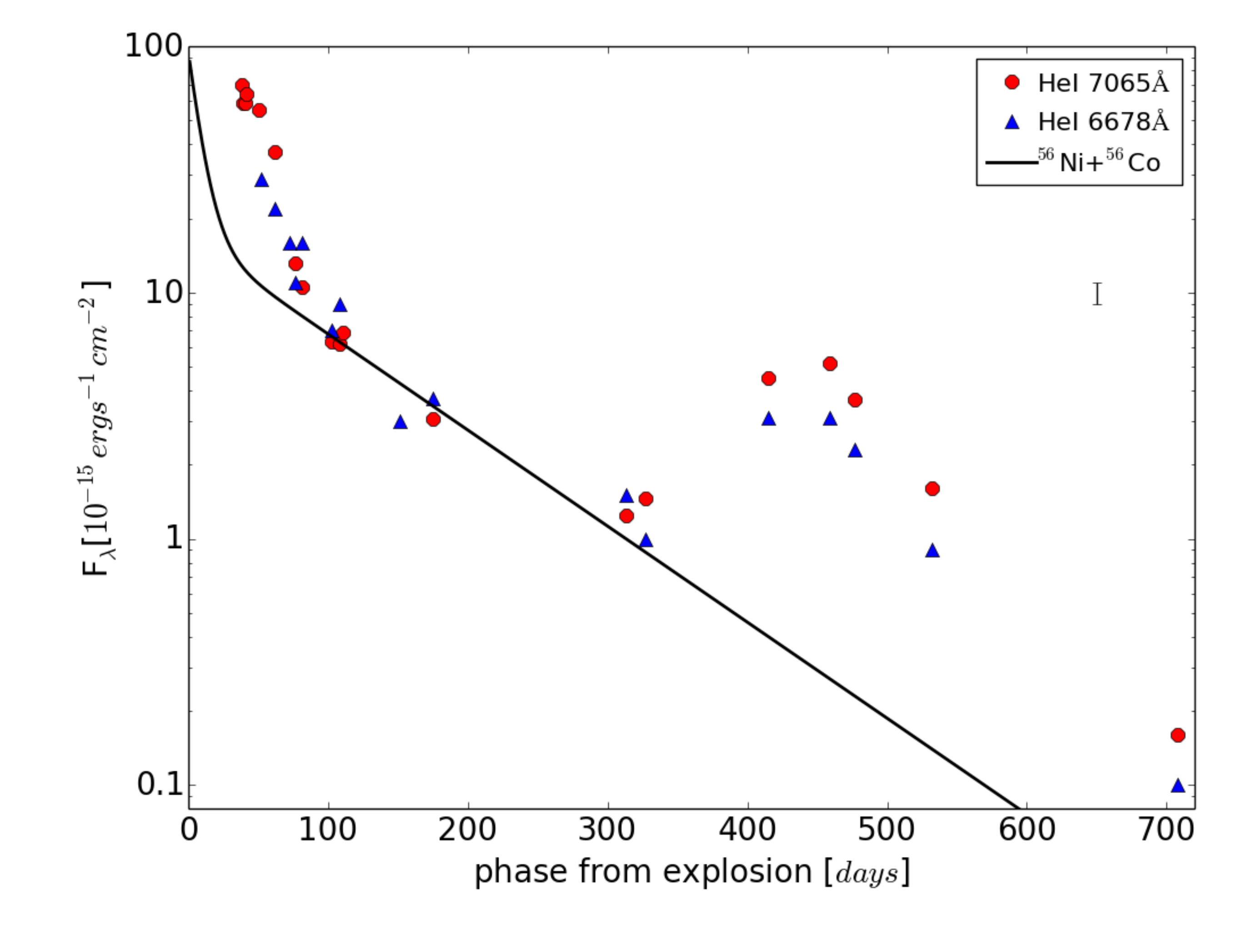}
\caption{Flux evolution of the He I 6678 \AA~ and 7065 \AA~lines. The fluxes have been reddening corrected. The $^{56}$Ni+$^{56}$Co fitting law is similar to that used for fitting the bolometric light curve of Figure \ref{bol_fig} and has been tuned to fit the $+100 \div +300$ days flux points. As in Figure \ref{bol_fig}, a rise time to maximum of 10 days has been assumed. A mean error-bar for the line fluxes is reported.}
\label{HeI_flux}
\end{figure}

Figure \ref{hei7065b} shows the evolution of the He 7065 \AA~emission up to almost two years after explosion. While the flux evolution of the line shows a bump between +300 and +500 days (see Table \ref{lines2} and Figure \ref{HeI_flux}), it becomes narrower with a terminal velocity of about 1200 \kms~(see Figure \ref{vel_evol}). 

The flux evolution of the He I 6678 \AA~ and 7065 \AA~lines reported in Figure \ref{HeI_flux} shows that the evolution of the line fluxes is consistent with that of the $^{56}$Co decay in the time interval $+100 \div  +300$ days, while at earlier and later phases some extra energy from the interaction between the ejecta and CSM is required.\\
Despite the CSM/ejecta interaction is likely the primary source for the SN luminosity, the mechanism responsible for the evolution of the  He\,I emission 
is far from being clear.
On day +449, the He\,I 7065\,\AA\ luminosity is $\sim 4\times10^{38}$\,erg\,s$^{-1}$.
The He\,I 5876\,\AA\ and He\,I 10830\,\AA\ lines are typically stronger than
   the He\,I 7065\,\AA\ line by 1.2 and $4-5$ times, respectively
   \citep{alm89}.
The total He\,I luminosity around day +450 is estimated to be 
   $\sim$2.8$\times10^{39}$\,erg\,s$^{-1}$, which is larger than the total
   H$\alpha$ luminosity of $2\times10^{39}$\,erg\,s$^{-1}$ and significantly 
   larger than the luminosity of the H$\alpha$ Core component.
This fact and the low velocity of He-emitting gas suggest, again, that the helium
   emission originates from the He-rich central zone of the SN ejecta.

Interestingly, the H$\alpha$ line does not show a significant luminosity enhancement during
   $+300 \div +450$ days (see Figure \ref{Ha_flux}), when the He\,I 7065\,\AA\ flux increases by a factor of 4.
This may be understood if the hydrogen ionisation fraction
   is close to unity, so that the hydrogen emission does not react to the enhanced
   ionisation that leads to the emergence of strong helium lines.
If this is the case, then the total mass of hydrogen in the SN ejecta should be
   close to the the mass of the ionized hydrogen, i.e., of the order of
   $0.1 - 0.3~M_{\odot}$ (see Sect. \ref{modelHa}). Alternatively, if the helium is ionised by the central
   source, the X-UV radiation could be strongly attenuated already in the He layer, so that the external hydrogen layer would be only marginally affected. 

Two possible sources for powering the transient He\,I emission are conceivable:
 X-ray emission from the enhanced CSM interaction or energy release due to accretion into the black hole.
The latter requires the accretion of $\sim$10$^{-7}$\,$M_{\odot}$ 
   during about +200 days and a super-Eddington X-UV luminosity.
This could originate in the thermalisation of the accretion
   disk wind in a scenario of disk outflow interacting with the SN ejecta
   \citep{dex13}.

\subsection{Photospheric radius}\label{phot_radius}
\begin{figure}
\includegraphics[width=9.0cm,angle=0]{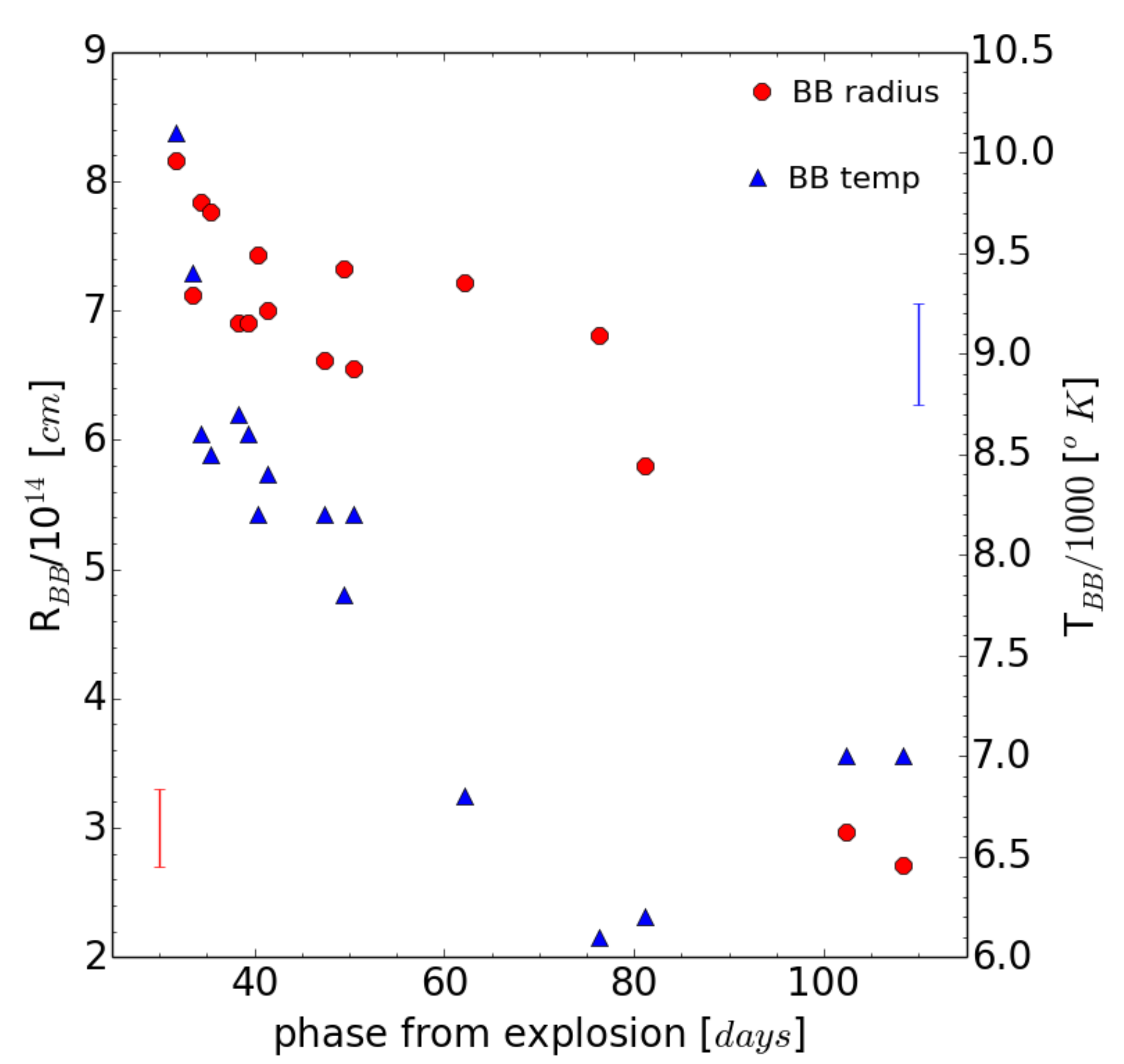}
\caption{Evolution of the blackbody radius and the temperature in days from the explosion, assumed to occur 10 days before the maximum light. The average temperature error bar of $\pm 250^o$K is shown in the upper-right corner, while the mean error bar for radius ($\pm 0.3\times 10^{14}$ cm) is shown in the lower-left corner.}
\label{ratemp}
\end{figure}

In order to derive the evolution of the photospheric radius and temperature, we perform a blackbody fit to our early time spectra, after correcting for reddening and redshift values reported in Table \ref{data}. The fits were done selecting the portions of the spectra free of evident lines (examples of the fits are reported in Figure \ref{first}). The fit errors, which depend on a number of parameters, mainly the wavelength extension of the spectra, lines contamination, the uncertainty of the flux calibration and the assumed extinction, have been estimated to be of the order of 250 $^o$K. Adopting the bolometric curve of Figure \ref{bol_fig}, spherical symmetry and a filling factor of 1, we estimate the evolution of the blackbody radius and the temperature (Figure \ref{ratemp}). The computation is extended up to 100 days post explosion. This does not imply that the SN has a clear spheric symmetric photosphere up to that epoch (cfr. next Section).

The photosphere several days after the explosion is relatively hot, with a T$_{BB} \sim 10000~^{\mathrm{o}}$K and a radius close to $7.4\times 10^{14}$ cm. While the temperature shows a smooth decrease to about $6500~^{\mathrm{o}}$K at a phase of +100 days after explosion, the photospheric radius remains almost at a constant value of about $6.9\times 10^{14}$ cm up to 80 days after explosion, with a sudden decline to $3.0\times 10^{14}$ cm at phase about $+100 \div +110$ days (see Figure \ref{ratemp}).

\subsection{Parameters of the progenitor star and environment geometry}\label{geo}
In the following, we will sketch a scenario for the SN explosion and its environment that will attempt to account for all available observations.\\
Despite a high expansion velocity of the ejecta ($\sim$10$^4$\,km\,s$^{-1}$, see Table \ref{lines}) as indicated by
   the broad wing of H$\alpha$ and the boxy He\,I 5876\,\AA\ emission,
   the spectra have never shown pronounced broad absorption lines of either
   hydrogen or other species.
This implies that the photosphere is located in the external layer
   of the SN ejecta.
The absence of pronounced broad absorptions is typical of SNe~IIn, in which the SN ejecta
   interact with a dense circumstellar matter (CSM), and is attributed to
   the high optical depth in the cool dense shell (CDS), that is located at the ejecta/CSM interface \citep{chu01}. 
It should be emphasised that - in the case of SN~1996al - the CDS is probably 
patchy because we can detect the broad H$\alpha$ emission
   arising in the inner SN ejecta. 
   
The higher resolution spectrum on day +43 (see Figure \ref{Ha_evol}) showing the pronounced broad and narrow
   H$\alpha$ components can be used to probe both the SN ejecta and the CSM.
The broad component is emitted by the freely expanding ejecta with  maximum
   velocity of $\la 10000$\,km\,s$^{-1}$, while the narrow line with P-Cygni profile presumably forms in the CSM, which expands with
   $v < 2000$\,km\,s$^{-1}$.\\
With these numbers, it is easy to calculate the radius of the CDS during the first 1-2 months after the explosion (assuming it occurred 10 days before the maximum light), which spans between $3-4\times 10^{15}$ cm. Given the radii calculated in the previous section, and supposing that the photosphere is placed close to the CDS, an upper limit of $\sim10$\% for the filling factor of the patchy CDS is derived.

\subsubsection{Modelling the early \Ha~profile}\label{modelHa}

The broad \Ha~line is modelled with the Monte Carlo technique adopting
   the recombination emission in the ionised ejecta, with the H$^+$ and $n_e$
   distributions described by a broken power law ($n_e \propto v^k$) with
    power index $k = k_1$ in the inner zone ($v<v_0$) and $k = k_2$ in
   the outer zone. 
The Thomson scattering with thermal broadening in the ejecta is taken into
   account, assuming a kinetic temperature of $10^{4}\,^{\mathrm{o}}$K.
Note that  Thomson scattering turns out to be a crucial diagnostic probe 
    for estimating the total amount of ionised hydrogen.    
Since our model does not compute the hydrogen level population,
we consider two extreme cases for the H$\alpha$ optical depth: 
   optically thin H$\alpha$ line ($\tau_{23} \ll 1$) and optically thick one
   ($\tau_{23} \gg 1$).
The optically thin case is plausible, because the flux in the Balmer continuum
   seems to be large enough to strongly depopulate the second level. 
Two sources of continuum  are included: the SN ejecta with the emissivity
   $\epsilon \propto n_e^2$, and the CDS at $v=10000$\,km\,s$^{-1}$. 
We do not compute the absolute flux neither in the line nor in the continuum.
We neglect the contribution of the line emission from the CDS compared 
to the continuum radiation bearing in mind that, in the case of 
the opaque CDS, the contrast of the CDS line emission with respect to 
the CDS continuum is small. This is indicated by observations of 
interacting supernovae \citep[e.g. SN~1998S at the age of 10-40 days, 
cf.][]{fas01} and supported by arguments based 
on the model of a thin opaque shell with the estimated line-to-continuum 
contrast of the order of the thermal-to-expansion velocity ratio \citep{chu01}.
Instead, we let line and continuum emissivities of SN ejecta to fit
the observations. The contribution of the CDS in the line emission is then
neglected whereas the continuum radiation of the CDS is taken into account. The angular dependence of the CDS continuum brightness ($I_c$)
affects the strength of the H$\alpha$ broad absorption. 
We assume $I_c \propto 1/\cos{\theta}$, where $\theta$ is the angle with the normal to the CDS surface. This approximation holds when the CDS is either optically thin or fragmented in such 
a way that the average number of clouds on the line of sight is less than unity.
Strong fragmentation is expected given the fact that CDS is Rayleigh-Taylor 
unstable \citep{che82}. The contribution of the CDS 
continuum is a free parameter derived from the best fit of the \Ha~profile. 
The latter, is determined by the normalised distribution of ionised hydrogen along the radius and by the Thomson optical depth.

\begin{figure}
\includegraphics[width=8.5cm,angle=0]{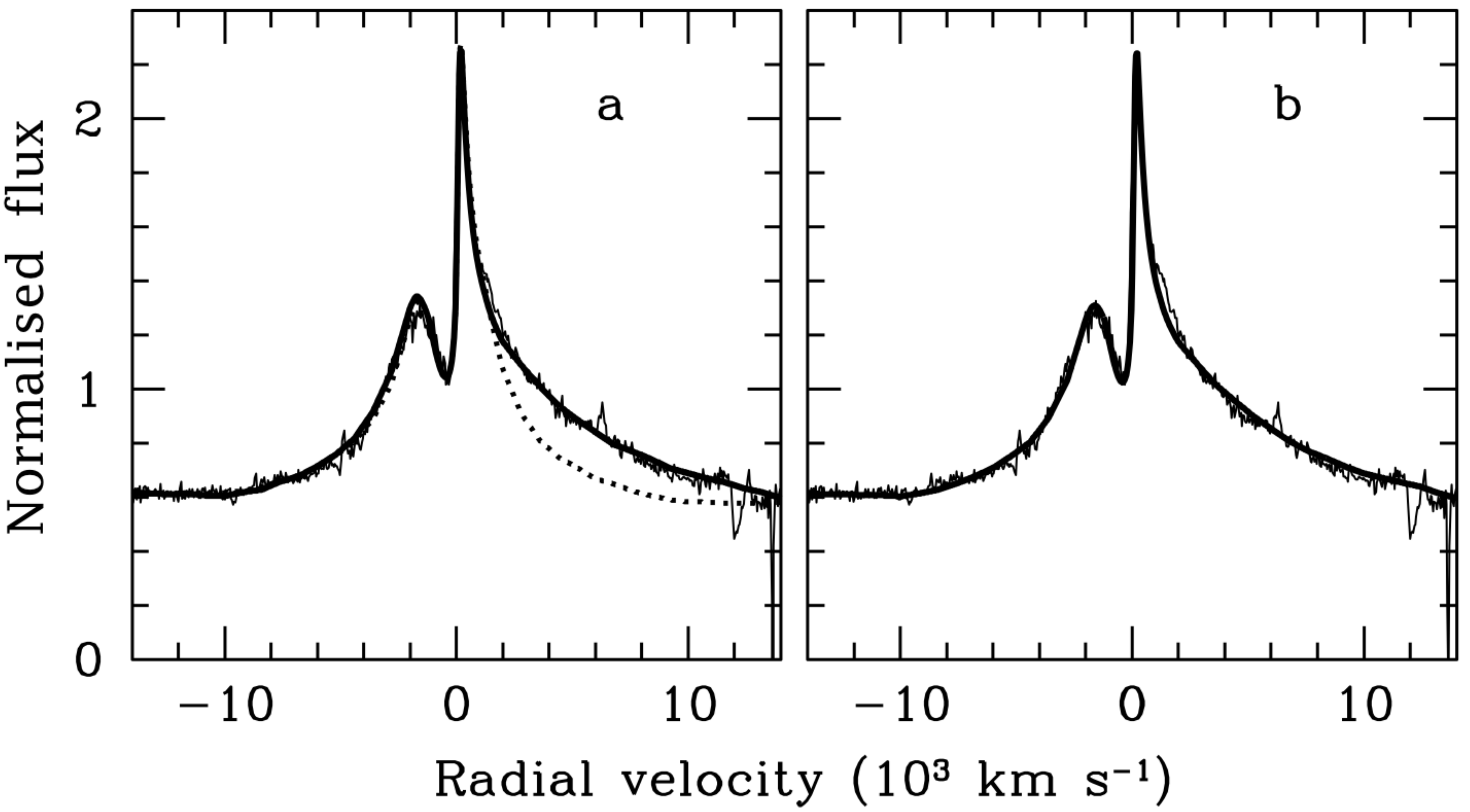}
\caption{Comparison of the H$\alpha$ line model with the observed spectrum on day +43
   ({\it thin solid} line).
{\it Panel a}: the SN ejecta with optically thin H$\alpha$ line
   ({\it thick solid} line) and the same model without Thomson scattering
   ({\it dotted} line). 
{\it Panel b}: model with optically thick H$\alpha$ line. 
In both cases the narrow component forms in the CSM and is primarily due to 
   resonance scattering.}
\label{fig:fhabn}
\end{figure}

We show two cases for the broad component: $\tau_{23} \ll 1$ (no H$\alpha$
   resonance scattering) and $\tau_{23} \gg 1$ (Figure~\ref{fig:fhabn}).
A reasonable fit of the broad component to the observed H$\alpha$ profile
   on day +43 is found for the parameters: $k_1 = 1$,  $k_2 = -1.7$,
   $v_0 = 1800$\,km\,s$^{-1}$, and Thomson optical depth $\tau_T = 2.1$.
The case without Thomson scattering (dotted line in Figure~\ref{fig:fhabn}a) demonstrates
   the crucial role of electron scattering for the formation of the line
   profile. 
In the case of the optically thick H$\alpha$ line, we find that the relative
   contribution of the ejecta continuum is $\leq10$\%; for a larger contribution of the ejecta continuum, the broad H$\alpha$ 
absorption component becomes prominent. 
For the optically thin H$\alpha$, the calculated recombination continuum in the Case B, can contribute only to 3.3\% of the observed continuum.
The recovered Thomson optical depth, together with the estimated CDS radius on day +43 and the normalised electron distribution recovered from the H$\alpha$ profile, suggest a mass of ionised hydrogen of
   0.07\,$M_{\odot}$ in the ejecta and a H$\alpha$ luminosity of
   $2\times10^{40}$\,erg\,s$^{-1}$ on day +43. 
The latter is in excellent agreement with the observational estimate. 

We now turn to the narrow line.
The peak of the narrow emission on days +23 and +43 suggests
   the presence of the H$\alpha$-emitting circumstellar gas with
   velocities of $\sim 300$ km s$^{-1}$. 
On the other hand, the wings of the narrow emission profile and the blue edge of
   the narrow absorption indicate CSM velocities as high as
   $\sim2000$\,km\,s$^{-1}$.
This is consistent with the expansion law $v \propto r$ for the CSM with a maximum
   velocity of $\sim2000$\,km\,s$^{-1}$ which is used for the H$\alpha$
   modeling.
We adopt the external radius for the fast CSM outflow of $5\times10^{15}$\,cm
   on day +43, keeping in mind that the spectra suggest that this kinematics
   is maintained at least until day +63. 
We find that the resonance scattering is a good approximation for the narrow
   P-Cygni line, although we need to add a net emission which amounts to 8\% of the 
   resonant scattering component.
We checked also a model where the CSM outflow has a constant velocity. However, such a model reproduced neither the absorption nor the emission components
   of the narrow line.
In the best-fit case shown in Figure \ref{fig:fhabn}, the H$\alpha$ optical depth in the 
CSM decreases outward from 0.9 at the CDS radius ($R_s$) down to zero at the 
boundary radius $R_b$ according to $\tau = 0.9[(r-R_s)/(R_b-R_s)]^{1.6}$. 

The kinematics $v\propto r$ followed by the observed narrow H$\alpha$ poses 
a question about its origin. Generally, this sort of the kinematics combined 
with large value of the terminal velocity ($\sim 2000$ km s$^{-1}$) suggests 
a mass ejection via shock wave. At present,
we are not able to propose a reliable mechanism for this vigorous 
mass loss. Beside the stellar shock, the common envelope phase in a 
binary system could also play a role.

\subsubsection{Modelling the light curve}\label{model_LC}

The fast initial ($t<+100$\,days) drop of the bolometric light curve (see Figure \ref{bol_fig}) strongly
   indicates that the diffusion time in the SN envelope is small, which
   implies low-mass ($\sim$1\,$M_{\odot}$) ejecta.
At first glance, this is at odds with the high mass of the SN progenitor suggested by
   the pre-explosion H$\alpha$ luminosity ($\sim$10$^{37}$\,erg\,s$^{-1}$, see Table \ref{progenitor}).
In fact, in the case B recombination, the pre-explosion H$\alpha$ requires a rate of ionising radiation of
   $\sim$10$^{49}$\,photons\,s$^{-1}$. 
To sustain this ionisation rate, we need a hot progenitor star with bolometric luminosity $\log{L/L_{\odot}}\ga5.4$ and radius $R\sim10R_{\odot}$ (see discussion in Sect. \ref{prog}). According to atmosphere models \citep{SS_08}, the above precursor may have been a $25\,M_{\odot}$ ZAMS star at the final evolutionary stage \citep{WHW_02}.
The apparent controversy between the SN light curve and pre-SN H$\alpha$ 
   constraints can be solved if one assumes that the $25\,M_{\odot}$ ZAMS
   progenitor had lost almost the whole hydrogen envelope, with most of the
   helium core falling back onto the collapsing core making a black hole \citep[see models in the $25-30$ \M~mass range of][]{heg03}. 

For our fiducial model we assume that the $25\,M_{\odot}$ progenitor
\citep{WHW_02} ends up as a stripped pre-SN star with a $8.3\,M_{\odot}$
helium core, and a $0.3\,M_{\odot}$ envelope,
with a hydrogen mass fraction of 0.5.
The adopted pre-SN radius is $R=10\,R_{\odot}$ constrained by the pre-explosion
H$\alpha$ luminosity.
To construct the stripped pre-SN configuration, we scaled the mass and the
radius of the hydrogen envelope of the $25\,M_{\odot}$ progenitor to
the required values.
Using a radiation hydrodynamics code \citep{utr07, UWJM_15}, we analysed the 
outcome of the explosion which is initiated by a supersonic piston applied
to the bottom of the stellar envelope at the boundary of the
2.1\,$M_{\odot}$ central core, which is then removed from the computational mass
domain and assumed to collapse into a neutron star.
We computed a set of SN models with  explosion energies of
 $(0.1-1.0)\times10^{51}$\,erg, ejecta masses of $0.45-2.1\,M_{\odot}$, and
kinetic energies of $(0.2-4.5)\times10^{50}$\,erg.
Note that the explosion energies released at the boundary of the collapsing
core are in the plausible range for the neutrino-driven explosion mechanism
\citep{Jan_12}.
The light curves of these models have  a very narrow shock breakout peak
($\sim$10$^{-4}$ days),  a short overall duration ($\sim$10 days), and a
 luminosity of $<$10$^{42}$\,erg\,s$^{-1}$ (after the shock breakout peak),
while the expansion velocity at the boundary is large
($>$10$^4$\,km\,s$^{-1}$). 
These properties are in agreement with the absence of broad absorption lines
in the spectra which, in turn, suggest a small energy contribution of the
intrinsic SN luminosity to the bolometric light curve, and instead support
the circumstellar interaction as the dominant source powering the SN luminosity.

The bolometric luminosity powered by the CS interaction is calculated in a
   thin shell approximation \citep{che82,chu01}.
The model suggests that freely expanding SN ejecta collide with dense
   CSM.
The interaction proceeds via forward and reverse shocks with the 
  CDS at the contact boundary between SN and CSM.
The CDS radius $R_s$ and velocity $v_s$ are recovered from the numerical
   solution of the equations of motion and mass conservation. 
The forward shock velocity is $v_s-v_w$ ($v_w$ is the wind velocity) and
   the reverse shock velocity is $R/t-v_s$.
The post-shock temperature is set by the shock velocity assuming an isothermal
   ($T_e=T_i$) strong shock, while the post-shock density is set to be four
   times the pre-shock density.
The X-ray luminosity of the $j$-th shock ($j=1$ for the forward shock and
   $j=2$ for the reverse shock) is $L_j=2\pi R_s^2 v_j^3\eta_j$, where $v_j$
   is the shock velocity, $\eta_j=t/(t+t_{c,j})$ is the cooling efficiency
   and $t_{c,j}$ is the post-shock cooling time.
The optical bolometric luminosity is calculated assuming that the full X-ray luminosity
    of both shocks is absorbed by the cool material of the CDS, the un-shocked
    SN ejecta and the CSM.

\begin{figure*}
\includegraphics[width=16cm,angle=0]{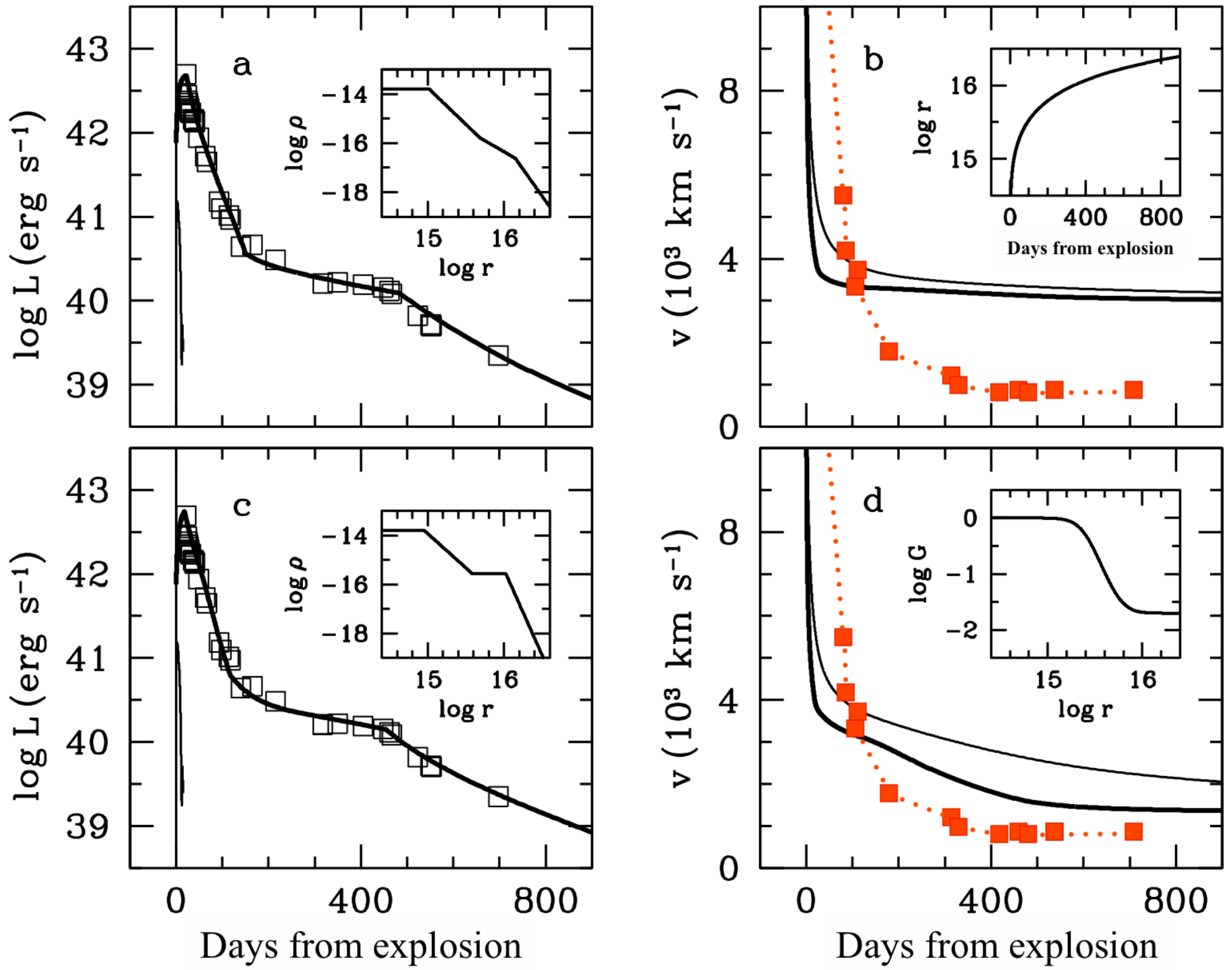}
\caption{
Bolometric light curve and expansion velocity in the CS interaction model. 
{\it Panel a}: light curves of the fast evolving, dim hydrodynamic model ({\it thin} line) 
   and the interaction model ({\it thick} line) in the spherical case
   compared to the observations ({\it open squares}).
The inset shows the CS density distribution, where $\rho$ and $r$ units are g\,cm$^{-3}$ and cm, respectively.
{\it Panel b}: the CDS velocity ({\it thick} line) and the boundary velocity
   of the un-shocked SN ejecta ({\it thin line}). For comparison, the observed velocities of the He\,I 7065 \AA~are also plotted ({\it solid, red squares} connected with a {\it dotted, red} line)
The inset shows the CDS radius.
{\it Panel c}: the same as in panel {\it a} but for the non-spherical case.
{\it Panel d}: the CDS velocity ({\it thick} line) and the boundary velocity
   of the un-shocked SN ejecta in the equatorial plane ({\it thin} line). For comparison, the observed velocities of the He\,I 7065 \AA~are also plotted ({\it solid, red squares} connected with a {\it dotted, red} line).
The inset shows the geometric factor describing the angular width of the equatorial
   disk (see text).
}
\label{fig:flc}
\end{figure*}
We consider two cases: 
   (i) spherically symmetric CSM with velocities $v=v_0(r/r_0)$ for $r<r_0$
   and $v=v_0=2000$\,km\,s$^{-1}$ for $r>r_0\approx 5\times10^{15}$\,cm and
   (ii) anisotropic CSM with a dense equatorial disk and a rarefied, more symmetric flow.
This latter outflow presumably does not contribute significantly to the interaction
   luminosity, although it is responsible for the narrow H$\alpha$ absorption.
The equatorial CSM is assumed to have a low expansion velocity: we set it
   to be 100\,km\,s$^{-1}$ (the result does not depend on the exact value of 
    $v_w$).
The CSM anisotropy is set as the fraction of a sphere occupied by
   the equatorial CSM, $G=\Omega(r)/4\pi$. This factor multiplied by
   the luminosity computed for the spherical CSM gives the luminosity
   in the anisotropic case.
Indeed, the luminosity and its temporal evolution constrain the CS density 
and its radial distribution. We find that the requirement of a rapid 
deceleration suggests a low mass and low energy supernova ejecta in order to minimise the ejecta momentum. A search for the best fit model in 
an extended parameter space leads us to conclude that one cannot describe 
simultaneously the light curve and the low expansion velocity in the 
framework of a spherically symmetric model. This is the primary 
reason to invoke an asymmetric CSM. We do not pursue the goal to find the 
best fit model for the CSM using a minimisation procedure. Our best-fit model 
should be considered rather as a reasonable possibility.

The model light curves and the expansion velocity of the CDS for the isotropic and anisotropic CSM cases are shown in Figure~\ref{fig:flc}. 
As parameters for the hydrodynamical model we adopt an ejecta mass
 $M=1.15\,M_{\odot}$ and a kinetic energy $E = 1.6\times10^{50}$\,erg.
The ejecta consist of $0.15\,M_{\odot}$ of hydrogen and nearly $1.0\,M_{\odot}$
of helium.
The rest of the He core is assumed to fall back, thus possibly making
a black hole with a mass of $7-8\,M_{\odot}$.
The SN explosion produces the brief shock breakout peak followed by a short
quasi-plateau ($\sim$12\,days) with a luminosity of $\sim$10$^{41}$\,erg\,s$^{-1}$.
   This contributes negligibly to the SN
   luminosity, which is instead dominated by CS interaction (Figures~\ref{fig:flc}, panels a and c). The modelled interaction luminosity for both the spherical and the anisotropic CSM well fits the
   observed light curve (Figures~\ref{fig:flc}, panels a and c), whereas the expansion
   velocities of the CDS at late times are quite different: the spherical model
   predicts deceleration down to 3000\,km\,s$^{-1}$, while the CDS for the
   equatorial CSM is decelerated down to 1300\,km\,s$^{-1}$
   (Figures~\ref{fig:flc}, panels b and d). 
In that respect, the non-spherical model is preferred, given the similarities of the model with the velocity evolution of \Ha~and the He\,I lines shown in Figure \ref{vel_evol}.
The total mass of the non-spherical CSM in the range of $r<3\times10^{16}$\,cm
   is $0.13\,M_{\odot}$.
The geometric factor is $G=0.02$ at the large radii (Figure~\ref{fig:flc}d,
   inset) which corresponds to a polar angle width of the equatorial disk
   of $23^{\circ}$.

The scenario with an equatorial CSM predicts a double-horn line emitted
   in the interaction zone.
Indeed, at $t>+142$\,days the H$\alpha$ shows Blue and Red bumps with radial
   velocities of about $-1100$\,km\,s$^{-1}$ and 1550\,km\,s$^{-1}$,
   respectively (see Table \ref{lines2}).
The velocity and intensity differences between those peaks indicate that
   the structure of the CSM deviates from axial symmetry.

\subsubsection{Evidence from  narrow absorption lines}\label{narrow_lines}
The narrow absorptions of the H$\alpha$, H$\beta$, and Fe\,II (mult. 42) lines
   are shallow in the higher resolution spectra of days +23 to +26 (see Sect. \ref{spec}).
Moreover, H$\alpha$ and H$\beta$ have similar relative depths, $A \approx 0.1$.
Given a theoretical ratio of optical depths H$\beta$/H$\alpha$ = 1/7,
   the facts that the H$\beta$ and H$\alpha$ absorptions have similar depths 
   and are both shallow imply that these lines are saturated, but only about
   10\% of the continuum source is covered by line absorbers. 
The spherical component of the CSM responsible for the narrow absorptions 
   is therefore clumpy and consists of clouds optically thick in the H$\alpha$
   and H$\beta$ lines embedded in the optically thin inter-cloud gas.
The interaction with the clumpy spherical CSM might contribute to the central (Core)
   H$\alpha$ emission component ($t > 142$ d).
The fact that this emission component is redshifted by $300-400$\,km\,s$^{-1}$
   could be attributed to the asymmetry of the CSM in a sense that the far side
   of the quasi-spherical CSM is more dense and mostly contributes to the
   Core H$\alpha$ emission. 

\subsubsection{The progenitor scenario}
The dominant role of the continuum originated by the CSM-ejecta interaction and
   the significant deceleration of the SN ejecta indicate lower ejecta
   mass and explosion energy than in normal core-collapse SNe.
We have shown in the previous Sect. that ejecta with $M_{ej} \approx 1.15~M_{\odot}$ and
   $E \approx 1.6\times10^{50}$\,erg are consistent with the observations.
The low hydrogen mass ($\sim 0.15~M_{\odot}$) and the high He\,I luminosity
around day +450 imply that the ejecta are made up primarily of helium ($\sim 0.8\,M_{\odot}$).

\begin{figure}
\includegraphics[width=9cm,angle=0]{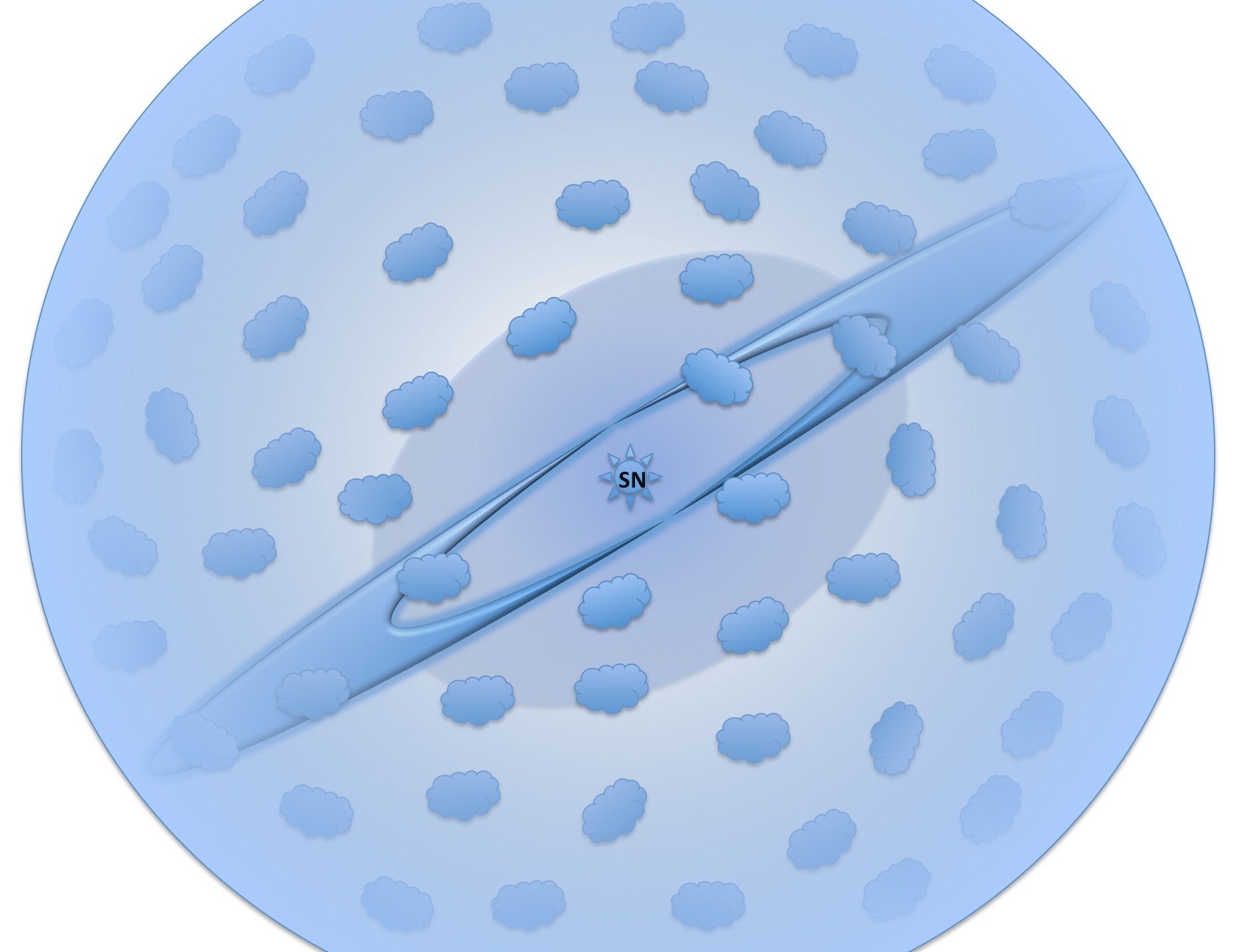}
\caption{Schematic cartoon of the CSM shape surrounding SN 1996al (not to scale). Line of sight is perpendicular to the figure plane. See text for a description.}
\label{CSM_shape}
\end{figure}

The CSM close to the exploding star has an overall spherical distribution (see Figure \ref{CSM_shape}), but with some density enhancement in an oblate structure to explain the asymmetric, early He\,I line profiles.
The inner radius of the equatorial disk is $r \approx 3\times10^{15}$\,cm, while its external limit is $> 5\times 10^{17}$ cm ($> 0.15$ pc), since the ejecta are still interacting with the disk at the time of our last observation (+5542 day). The disk is embedded in a more spherically symmetric, clumpy CSM.
The mass of the CSM in the anisotropic model, including the inner mostly spherical zone is 0.13 \M, within a distance $r<3\times10^{16}$\,cm. This value is very similar to the amount of CS mass found in the twin SN~1994aj (Paper I) and about ten times more than in SN~1996L (Paper II).
According to this scenario, and given the progressive quenching of the \Ha~ Core and Red components, we may suppose that both the receding part of the disk and the quasi-symmetric, clumpy CSM get shallower and shallower in density with time, or that some obscuring dust is forming mostly inside the shocked ejecta (or a combination of the two).

The presence of dense CSM around the progenitor star of SN~1996al is also supported by the recovery of the progenitor in an archive \Ha~image, from which a \Ha~luminosity of logL$_{H\alpha} \sim 37.28$ is derived. This \Ha~luminosity implies a massive $\sim 25\,M_{\odot}$ ZAMS progenitor star (see Sect. \ref{prog}), which has lost almost the entire hydrogen envelope during its evolution.

It is interesting to note that such a highly asymmetric geometric configuration derived for the SN 1996al CSM has been observed around very massive evolved stars such as MN18 of Large Magellanic Cloud (LMC), and in other blue supergiants (including LBV candidates) of LMC and Milky Way, including the SN 1987A's progenitor \citep{gva15}.

\begin{figure}
\includegraphics[width=9cm,angle=0]{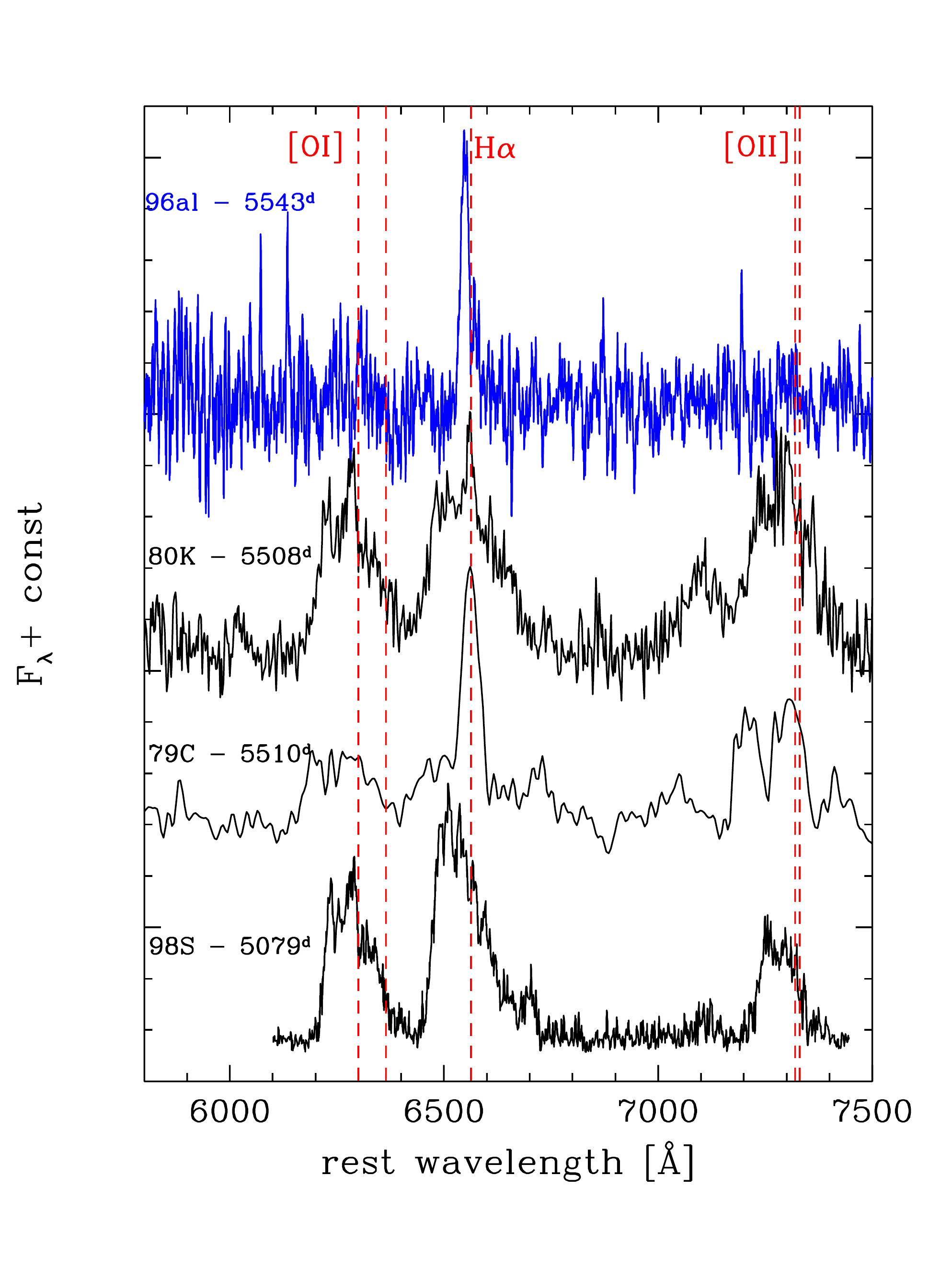}
\caption{Comparison of spectra taken about 14 years after explosion of our SNe sample. The SNe 1979C and 1980K spectra are from \citet{fes99}, while that of SN 1998S is from \citet{mau12}. }
\label{late}
\end{figure}

In this scenario, the absence of [O I] 6300-6364 \AA~emission lines at any phase (see Figure \ref{late}) is naturally explained since most of the oxygen layer has fallen onto the black hole\footnote{It is interesting to note that a similar scenario (low energy, $4 \times 10^{50}$ erg; massive, 26 \M, progenitor; substantial fallback of material on to the collapsed core) has been proposed to explain the physical characteristics of SN 1997D \citep{tur98,ben01}. However, this high mass scenario has been questioned by \citet{chu00} who, instead, proposed a low mass progenitor for SN~1997D. In the case of SN~1996al the key parameter that point towards a high mass progenitor scenario is the pre-explosion \Ha~luminosity.}.

\section{Conclusions}\label{conc}
Extensive photometric and spectroscopic observations of SN~1996al along 15 years of its evolution are presented. The light curve indicates that the SN is of Type IIL, reaching peak absolute magnitudes of $M_B \sim -18.6$ and $M_V \sim -18.2$. However, a closer inspection shows that the decline rate changes with time, and this is correlated with significant changes in the spectral appearance. The complex SN evolution is very likely a consequence of the primary role played by the ejecta-CSM interaction during all phases of the SN life.

The spectra obtained soon after the maximum light show narrow P-Cygni Balmer lines superimposed on broader components with asymmetric profiles. In addition, broad and asymmetric He I lines in emission are detected. At these early phases, the broad H component and the He lines form in the same outer region of the ejecta. 
By day $+142$, the \Ha~profile dramatically changes: the narrow emission component disappears, and \Ha\/ splits in three components of comparable widths, that remain at constant wavelengths afterward.  
This composite profile remains visible over 15 years, although the relative strengths of the three components change with time.

The He lines behave differently. After day $+100$, they become significantly narrower ($\le1000$ \kms)  and symmetric, a clear indication that they form in a different region of the ejecta.
Later on, between day +300 and +600, they experience a sudden increase in flux. Then, the flux of the He~I line decreases again, and by day $+700$, these lines fade below the detection threshold. 

We have also recovered information on the progenitor star by inspecting a pre-explosion archive image in the \Ha~narrow band. A luminous source is detected at the location of SN~1996al,  showing a total \Ha~luminosity of 1.9 $\times$ 10$^{37}$ erg s$^{-1}$, which favours a massive ($25\,M_{\odot}$ ZAMS) progenitor star for SN~1996al. This, along with the SN parameters derived by the modelling, suggests that the progenitor was stripped of a large fraction of the hydrogen envelope in the latest stages of its life which ended with the collapse of a $7-8$ \M~ helium core.

The peculiar evolution of the H and He I lines is likely a consequence of the geometry, the non-uniform density profile of the CSM surrounding SN~1996al and the fallback onto the compact remnant. Observational evidences favour an highly structured CS scenario characterised by inner, high-density material with oblate structure plus an equatorial ring extended over 0.15 pc, both embedded in a more spherically symmetric, lower-density but clumpy CSM. The total mass of the CSM is modest, being $\ga 0.13$ \M~ within a distance $\la3\times10^{16}$\,cm.

Models show that the luminosity of  SN~1996al is best sustained by SN ejecta-CSM interaction. 
The inferred ejected mass is relatively small ($\sim 1.15$ \M, $\sim 0.15~M_{\odot}$ of which is hydrogen, the residual fraction is helium) and is expelled in a low kinetic energy explosion ($\sim 1.6\times10^{50}$\,erg).
The amount of ejected $^{56}$Ni (if any) expelled in the explosion is constrained to be $\la$ 0.018\M.
The modest ejecta and $^{56}$Ni mass are explained with a massive fallback of material into the compact remnant, producing a $7-8\,M_{\odot}$ black hole.

\bigskip
\noindent
{\bf ACKNOWLEDGMENTS}

\noindent
We thank the anonymous referee for very useful comments and suggestions.\\
We thank T. Augusteijn, M. Birlan, J. F. Le Borgne, S. Covino, P. Rosati and G. C. Van de Steene for giving us part of their telescope time; P. Fouqu\'e for supplying the DENIS photometry; and M. Riello for performing some of the observations.\\
We are in dept with N. Richardson for providing unpublished spectra of LBVs.\\
We thank R. McCray and J. Marcaide for helpful discussions and comments on the manuscript.\\
SB, MT, AP are partially supported by the PRIN-INAF 2014 project Transient Universe: unveiling new types of stellar explosions with PESSTO. VPU is supported by Russian Scientific Foundation grant 14-12-00203. NC thanks the Program of RAS "Explosions in Astrophysics" for financial support. G.P. acknowledge support provided by the Millennium Institute of Astrophysics (MAS) through grant IC120009 of the Programa Iniciativa Cientifica Milenio del Ministerio de Economia, Fomento y Turismo de Chile.\\
This work is based on observations collected at the ESO-La Silla (under programs ESO N$^{\rm o}$ 151.D-0004, 56.D-0478, 57.D-0534, 58.D-0307, 59.D-0332, 60.D-0415, 61.D-0630, 57.E-0646); and ESO-Paranal (under programs ESO N$^{\rm o}$ 67.D-0422, 073.D-0670, 080.D-0213, 084.D-0265, 087.D-0693) Observatories.\\
One spectrum has been collected at the Anglo Australian Observatory by R. Stathakis, B. Schmidt and S. Woodings and retrieved via AAT Archive (http://apm5.ast.cam.ac.uk/arc-bin/wdb/aat\_database/observation\_log/make).\\
This research has made use of the NASA/IPAC Extragalactic Database (NED) which is operated by the Jet Propulsion Laboratory, California
Institute of Technology, under contract with the National Aeronautics and Space Administration. We have also made use of the Lyon-Meudon Extragalactic Database (LEDA), supplied by the LEDA team at the Centre de Recherche Astronomique de Lyon, Observatoire de Lyon.


\newpage

\appendix
\section{}
\subsection{Light curves and spectroscopic log of SN 1996al}

\begin{table*}
\caption{Optical and NIR photometric measurements for SN~1996al.}\label{obs_tab}
\begin{flushleft}
\begin{tabular}{lccccccccclc}
\hline   
 date  &MJD   &ph&   $U$    &  $B$      &     $V$   &    $R$    &  $I$      &  $J$      &  $H$         &        tel. & seeing\\
       &         &days$^{\dag}$&         &         &         &         &         &         &         &             & arcsec \\
23/7/96 &50287.29&22.8&14.27(08)&14.80(05)&14.61(04)&14.37(04)&14.25(04)&         &         &          Dut &1.3\\
25/7/96 &50289.32&24.8&14.54(07)&14.98(09)&14.77(03)&14.48(03)&14.36(03)&         &         &          Dut &2.0\\
26/7/96 &50290.34&25.8&14.68(05)&15.05(04)&14.81(03)&14.56(03)&14.42(03)&         &         &          Dut &1.7\\
27/7/96 &50291.35&26.9&14.76(05)&15.13(04)&14.90(03)&14.62(03)&14.49(03)&         &         &          Dut &2.4\\
28/7/96 &50292.32&27.8&14.88(05)&15.19(04)&14.96(03)&14.69(03)&14.56(02)&         &         &          Dut &1.7\\
29/7/96 &50293.33&28.8&14.96(04)&15.26(03)&15.01(04)&14.72(03)&14.60(04)&         &         &          Dut &1.6\\
29/7/96 &50293.40&28.9&         &         &         &         &         &14.27(05)&         &          Den &---   \\
30/7/96 &50294.40&29.9&15.03(09)&15.34(05)&15.03(02)&14.75(04)&14.62(02)&         &         &          Dut &1.4\\
1/8/96  &50296.31&31.8&15.00(08)&15.36(05)&15.11(02)&14.81(03)&14.70(02)&         &         &          Dut &2.2\\
6/8/96  &50301.30&36.8&15.31(09)&15.55(04)&15.25(03)&14.95(03)&14.83(03)&         &         &          Dut &1.5\\
7/8/96  &50302.28&37.8&15.32(07)&15.57(04)&15.28(03)&14.94(03)&14.82(03)&         &         &          Dut &2.3\\
7/8/96  &50302.43&37.9&         &         &15.29(03)&         &         &         &         &          EF1 &1.5\\
8/8/96  &50303.43&38.9&         &         &15.26(02)&         &         &         &         &          EF1 &2.4\\
9/8/96  &50304.43&39.9&         &         &15.27(03)&         &         &         &         &          EF1 &1.9\\
10/8/96 &50305.27&40.8&         &         &15.27(03)&         &14.83(04)&         &         &          Dut &1.4\\
10/8/96 &50305.43&40.9&         &         &15.31(03)&14.95(02)&         &         &         &          EF1 &1.4\\
18/8/96 &50313.35&48.9&16.09(05)&16.22(03)&15.79(03)&15.41(03)&15.29(02)&         &         &          Dut &2.3\\
1/9/96  &50327.37&62.9&         &         &16.30(03)&         &         &         &         &          EF1 &1.5\\
5/9/96  &50331.33&66.8&17.11(06)&17.14(05)&16.48(03)&16.02(04)&15.86(03)&         &         &          EF2 &2.1\\
1/10/96 &50357.27&92.8&         &18.33(05)&17.76(04)&17.29(03)&17.13(03)&         &         &          DF  &1.5\\
8/10/96 &50363.12&98.6&18.34(06)&18.50(05)&17.99(03)&17.50(03)&17.35(03)&         &         &          EF2 &2.5\\
24/10/96&50379.06&114.6&18.63(08)&18.65(07)&18.12(04)&17.64(05)&17.47(04)&        &         &          Dut &2.5\\
28/10/96&50383.13&118.6&18.71(07)&18.78(05)&18.24(04)&17.75(04)&17.60(04)&        &         &          Dut &1.4\\
18/11/96&50405.02&140.5&         &         &19.22(04)&         &         &        &         &          EF1 &1.4\\
18/11/96&50405.08&140.6&19.44(06)&19.66(05)&19.24(04)&18.46(04)&18.36(04)&        &         &          Dut &1.5\\
1/12/96(*) &50418.15&153.6&         &         &         &         &         &18.45(20)&18.33(20)&     IR2 &0.7\\
13/12/96&50430.00&165.5&         &19.56(05)&18.97(03)&18.39(03)&18.25(03)&        &         &          EF2 &2.0\\
1/2/97  &50480.05&215.6&         &         &19.68(05)&18.85(04)&         &        &         &          Dut &1.5\\
13/5/97 &50581.41&316.9&20.92(09)&20.85(05)&20.39(05)&19.43(05)&19.69(04)&        &         &          EF2 &1.5\\
14/5/97 &50582.35&317.9&         &         &20.41(06)&19.46(06)&         &        &         &          EF2 &1.4\\
15/6/97 &50614.54&350.0&         &         &20.20(06)&19.34(06)&         &        &         &          DF  &1.8\\
9/8/97  &50669.40&404.9&         &20.71(09)&20.26(08)&19.46(06)&19.83(06)&        &         &          DF  &1.7\\
16/9/97 &50708.00&443.5&         &         &         &         &         &19.05(20)&19.11(20)&         IR2 &1.2\\
17/9/97 &50709.00&444.5&         &         &         &         &         &18.92(20)&19.08(20)&         IR2 &1.1\\
22/9/97 &50713.17&448.7&21.01(20)&20.80(10)&20.43(08)&19.52(09)&19.92(08)&        &         &          EF2 &1.4\\
23/9/97 &50714.13&449.6&         &         &         &19.66(10)&         &        &         &          EF2 &1.3\\
6/10/97 &50726.07&461.6&         &20.75(10)&20.54(09)&19.72(08)&20.05(08)&        &         &          Dut &1.3\\
12/10/97&50732.02&467.5&         &         &20.66(08)&19.74(07)&         &        &         &          DF  &1.7\\
17/11/97&50769.70&505.2&         &         &         &         &         &19.54(30)&        &          IR2 &0.9\\
5/12/97 &50787.08&522.6&         &21.59(10)&21.38(07)&20.31(07)&20.69(08)&        &         &          DF  &1.4\\
2/1/98  &50815.08&550.6&         &         &21.58(20)&20.59(10)&20.90(10)&        &         &          Dut &1.5\\
3/1/98  &50816.18&551.6&         &         &21.56(10)&         &         &        &         &          Dut &1.6\\
4/1/98  &50817.10&552.6&         &         &21.67(15)&20.54(10)&         &        &         &          Dut &1.5\\
5/1/98  &50818.05&553.6&         &         &         &         &         &20.40(40)&        &          IR2 &0.7\\
29/5/98 &50962.42&697.9&         &22.83(20)&22.51(10)&21.38(10)&         &        &         &          EF2a&1.4\\
30/5/98 &50963.40&698.9&         &         &         &21.39(10)&         &        &         &          EF2a&1.0\\
5/7/98  &50999.41&734.9&         &         &         &21.48(10)&         &        &         &          Dut &1.8\\
12/9/98 &51068.80&804.3&         &         &         &21.99(10)&         &        &         &          EF2a&1.1\\
\hline
\end{tabular}

(${\dag}$) epoch relative to $V_{max}$ occurred on MJD=50265\\

(*) for this epoch we give also $K$=17.68(20)\\

Dut = Dutch+CCD camera; pixel scale=0.44''/px\\
Den = ESO1.0m+DENIS; pixel scale=3''/px\\
EF1 = ESO3.6m+EFOSC1; pixel scale =0.61''/px\\
EF2 = MPI2.2m+EFOSC2; pixel scale =0.34''/px\\
DF  = D1.54m+DFOSC; pixel scale =0.39''/px\\
IR2 = MPI2.2m+IRAC2; pixel scale =0.28''/px\\
EF2a= ESO3.6m+EFOSC2; pixel scale =0.157''/px\\

\end{flushleft}
\end{table*}

\newpage

\begin{table}
\caption{Spectroscopic observations of SN~1996al.} \label{spec_tab}
\begin{tabular}{rcrclr}
\hline
\hline
date   & MJD    & phase$^*$  &   range   & tel.$^{**}$      & res.\\
       &        & (days)     &   (\AA)   &                  &(\AA)\\
\hline	    		                    
23/7/96 &50287.54& 23        & 4400-7000 & B\&C             &  4  \\
24/7/96 &50288.39& 24        & 4400-7000 & B\&C             &  4  \\
25/7/96 &50289.34& 25        & 4400-7000 & B\&C             &  4  \\
26/7/96 &50290.34& 26        & 4400-7000 & B\&C             &  4  \\
29/7/96 &50293.35& 29        & 4030-9720 & B\&C             &  9  \\
30/7/96 &50294.34& 30        & 4030-9720 & B\&C             &  9  \\
31/7/96 &50295.35& 31        & 4030-9720 & B\&C             &  9  \\
1/8/96  &50296.38& 32        & 4030-9720 & B\&C             &  9  \\
7/8/96  &50302.44& 38        & 3740-6930 & EF1              & 16  \\
8/8/96  &50303.44& 39        & 3740-6930 & EF1              & 16  \\
9/8/96  &50304.43& 40        & 3740-6930 & EF1              & 16  \\
10/8/96 &50305.43& 41        & 3740-9890 & EF1              & 16  \\
12/8/96 &50307.41& 43        & 4940-6930 & B\&C             &  3  \\
22/8/96 &50317.16& 53        & 3490-11080& B\&C             & 13  \\
1/9/96  &50327.38& 63        & 3730-6910 & EF1              & 16  \\
5/9/96  &50331.38& 67        & 3200-9250 & EF2              & 10  \\
10/9/96 &50336.15& 72        & 3520-10320& B\&C             & 13  \\
1/10/96 &50357.29& 93        & 3300-9050 & DF               & 11  \\
7/10/96 &50363.30& 99        & 5160-9270 & EF2              & 11  \\
7/10/96 &50363.30& 99        & 3430-10080& EF2              & 25  \\
9/10/96 & 50365.51& 101     & 5430-8890  & RGO             & 8 \\
19/11/96&50406.03& 142       & 3730-6910 & EF1              & 16  \\
13/12/96&50430.12& 166       & 3840-7990 & EF2              & 12  \\
30/4/97 &50568.40& 304       & 3600-9600 & B\&C             & 12  \\
14/5/97 &50582.35& 318       & 5220-9250 & EF2              & 11  \\
10/8/97 &50670.19& 406       & 5100-10170& DF               & 12  \\
22/9/97 &50713.73& 449       & 3500-9300 & EF2              & 10  \\
8/10/97 &50729.07& 465       & 3350-10040& B\&C             & 12  \\
11/10/97&50732.07& 468       & 3400-8740 & DF               & 12  \\
5/12/97 &50787.12& 523       & 3500-8990 & DF               & 11  \\
29/5/98 &50962.91& 698       & 3380-7500 & EF2a             & 18  \\
12/9/98 &51068.33& 804       & 3320-7500 & EF2a             & 18  \\
27/7/01 &52117.28& 1853      & 4180-8180 & FOR1             & 14  \\
16/6/02 &52441.31& 2177      & 4090-9630 & FOR1             & 10  \\
23/7/04 &53209.30& 2944      & 4220-9630 & FOR2             & 10  \\
08/12/07&54442.04& 4178      & 3900-9600 & FOR2             & 10  \\
11/10/09&55115.04&  4850     & 3500-24000& XS          & 1,0.8,2.8 \\
3/9/11&55807.07&     5542      & 3500-24000& XS          & 1,0.8,2.8 \\
\hline
\end{tabular}

* - relative to the estimated epoch of $V$ maximum (MJD=50265)

** - See note to Table~\ref{obs_tab} for telescope coding plus:\\
B\&C = ESO1.52m+B\&C \\
RGO = AAT+RGO \\
FOR1 = ESO-VLT+FORS1\\
FOR2 = ESO-VLT+FORS2\\
XS = ESO-VLT+XShooter\\
The resolutions of XS refer to the 3 arms at 5000, 6000 and 15000 \AA, respectively\\

\end{table}

\subsection{Spectral line parameters of SN 1996al}

\begin{table*}
\caption{Spectral lines parameters as derived from spectra of SN~1996al up to 100 days.} \label{lines}
\begin{flushleft}
\begin{tabular}{ccccccc}
\hline
\hline
 phase&                                              &                        &H$\alpha$ &                &He\,I 6678 & He\,I 7065 \\
(days)&                                               &                        &                 &                &               &                \\
                                        \cline{3-5}\\
        &                                                  &(Em)$_b$     &(Ab)$_n$    & (Em)$_n$&              &                \\
\hline
       & $\lambda_c$ (\AA)                    & 6564            & 6556          &   6566      &              &                \\
+23 & FWHM;V$_{I_0}^{blue}$ (km/s)&  6500;12200& 1050          &    400       &              &                \\
       & $\phi$$^*$                                 &  372             &  -35            &     93        &             &                \\
\cline{1-2}\\
       & $\lambda_c$ (\AA)                    & 6568            & 6557          &   6566      &              &                \\
+24 & FWHM;V$_{I_0}^{blue}$ (km/s)&  6600;12700& 930            &    390       &              &                \\
       & $\phi$$^*$                                 &  477             &  -41            &    107       &             &                \\
\cline{1-2}\\
       & $\lambda_c$ (\AA)                    & 6567            & 6556          &   6565      &              &                \\
+25 & FWHM;V$_{I_0}^{blue}$ (km/s)&  6600;12800& 970            &    380       &              &                \\
       & $\phi$$^*$                                 &  482             &  -43            &    103       &             &                \\
\cline{1-2}\\
       & $\lambda_c$ (\AA)                    & 6566            & 6556          &   6566      &              &                \\
+26 & FWHM;V$_{I_0}^{blue}$ (km/s)&  7000;12700&  990           &    410       &              &                \\
       & $\phi$$^*$                                 &  513             &  -40            &     99        &             &                \\
\cline{1-2}\\
       & $\lambda_c$ (\AA)                    & 6566            & 6551          &   6565      &              &  7065              \\
+29 & FWHM;V$_{I_0}^{blue}$ (km/s)&  6800;10300& 1030          &    290       &              & 10600               \\
       & $\phi$$^*$                                 &  560             &  -29            &     75        &             &     70           \\
\cline{1-2}\\
       & $\lambda_c$ (\AA)                    & 6565            & 6551          &   6565      &              &  7065              \\
+30 & FWHM;V$_{I_0}^{blue}$ (km/s)&  6900;12600& 1020          &    350       &              & 10800               \\
       & $\phi$$^*$                                 &  546             &  -30            &     73        &             &     59           \\
\cline{1-2}\\
       & $\lambda_c$ (\AA)                    & 6565            & 6551          &   6565      &              &  7065              \\
+31 & FWHM;V$_{I_0}^{blue}$ (km/s)&  6400;13100&  970           &    310       &              &  9600               \\
       & $\phi$$^*$                                 &  446             &  -25            &     54        &             &     59           \\
\cline{1-2}\\
       & $\lambda_c$ (\AA)                    & 6565            & 6550          &   6565      &              &  7065              \\
+32 & FWHM;V$_{I_0}^{blue}$ (km/s)&  6500;11900&  1020         &    300       &              &  10200               \\
       & $\phi$$^*$                                 &  482             &  -27            &     54        &             &     64           \\
\cline{1-2}\\
       & $\lambda_c$ (\AA)                    & 6572            & 6554          &   6565      &              &                     \\
+38 & FWHM;V$_{I_0}^{blue}$ (km/s)&  6300;12200&  550           &    450       &              &                     \\
       & $\phi$$^*$                                 &  503             &  -32            &     71        &             &                      \\
\cline{1-2}\\
       & $\lambda_c$ (\AA)                    & 6567            & 6553          &   6566      &              &                     \\
+39 & FWHM;V$_{I_0}^{blue}$ (km/s)&  5400;11500&  940           &    680       &              &                     \\
       & $\phi$$^*$                                 &  471             &  -37            &     41        &             &                      \\
\cline{1-2}\\
       & $\lambda_c$ (\AA)                    & 6565            & 6548          &   6563      &              &      7065               \\
+40 & FWHM;V$_{I_0}^{blue}$ (km/s)&  5800;11200&  580           &    620       &              &   10600                  \\
       & $\phi$$^*$                                 &  523             &  -26            &     39        &             &       55               \\
\cline{1-2}\\
       & $\lambda_c$ (\AA)                    & 6568            & 6549          &   6567      &              &                     \\
+41 & FWHM;V$_{I_0}^{blue}$ (km/s)&  5600;12000&  530           &    680       &              &                     \\
       & $\phi$$^*$                                 &  511             &  -21            &     33        &             &                      \\
\cline{1-2}\\
       & $\lambda_c$ (\AA)                    & 6569            & 6557          &   6566      &   6699           &                     \\
+43 & FWHM;V$_{I_0}^{blue}$ (km/s)&  5700;11800&  930           &    470       &   5100           &                     \\
       & $\phi$$^*$                                 &  440             &  -36            &     51        &     29        &                      \\
\cline{1-2}\\
       & $\lambda_c$ (\AA)                    & 6568            & 6555          &   6565      &   6712           &  7035                   \\
+53 & FWHM;V$_{I_0}^{blue}$ (km/s)&  5700;10200&  870           &    860       &   3600           &   10200                  \\
       & $\phi$$^*$                                 &  209             &  -30            &     66        &        22         &       37               \\
 \cline{1-2}\\
       & $\lambda_c$ (\AA)                    & 6570            & 6556          &   6565      &    6699          &                     \\
+63 & FWHM;V$_{I_0}^{blue}$ (km/s)&  5050;11000&  810           &    940       &     3300         &                     \\
       & $\phi$$^*$                                 &  169             &  -28            &     63        &       16          &                      \\
      
\hline
\end{tabular}

\end{flushleft}
\end{table*}

\begin{table*}
\contcaption{Spectral lines parameters as derived from spectra of SN~1996al up to 100 days.}
\begin{flushleft}
\begin{tabular}{ccccccc}
\hline
\hline
 phase&                                              &                        &H$\alpha$ &                &He\,I 6678 & He\,I 7065 \\
(days)&                                               &                        &                 &                &               &                \\
                                        \cline{3-5}\\
        &                                                  &(Em)$_b$     &(Ab)$_n$    & (Em)$_n$&              &                \\
\hline
       & $\lambda_c$ (\AA)                    & 6570            & 6557          &   6566      &    6700          &   7067                  \\
+67 & FWHM;V$_{I_0}^{blue}$ (km/s)&  4700;11100&  850           &    650       &     3250         &    5500                 \\
       & $\phi$$^*$                                 &  148             &  -23            &     38        &       11          &     13                 \\
\cline{1-2}\\
       & $\lambda_c$ (\AA)                    & 6568            & 6558          &   6567      &    6687          &   7059                  \\
+72 & FWHM;V$_{I_0}^{blue}$ (km/s)&  4400;10200&  1000         &    560       &     3700         &    4200                 \\
       & $\phi$$^*$                                 &  123             &  -18            &     28        &       16          &     11                 \\
\cline{1-2}\\
       & $\lambda_c$ (\AA)                    & 6563            & 6560          &   6566      &    6673          &   7065                  \\
+93 & FWHM;V$_{I_0}^{blue}$ (km/s)&  3500;8700 &  430            &    230       &     3300         &    3300                 \\
       & $\phi$$^*$                                 &  82              &  -10             &     11        &        7            &      6                 \\
\cline{1-2}\\
       & $\lambda_c$ (\AA)                    & 6562            & 6557          &   6565      &    6667          &   7061                  \\
+99 & FWHM;V$_{I_0}^{blue}$ (km/s)&  3200;10400&   570          &    460       &     3300         &    3700                 \\
       & $\phi$$^*$                                 &  85              &  -8               &      6         &        9            &      6                 \\
       \cline{1-2}\\
       & $\lambda_c$ (\AA)                    & 6562            & 6554          &   6564      &    6673          &                     \\
+101 & FWHM;V$_{I_0}^{blue}$ (km/s)&  3250;10600&   200:          &    450       &     3600         &                     \\
       & $\phi$$^*$                                &  90               &  -7               &      3         &        95            &                       \\      
\hline
\end{tabular}

$^*$ In units of $\times 10^{-15}$ erg\,s$^{-1}$\,cm$^{-2}$; not corrected for reddening.\\
\Ha~broad component better fitted with lorenztians up to phase +43d; after this phase they are better fitted with Gaussians. He\,I lines always fitted with Gaussians.\\
The velocities have been de-convolved for spectral resolution of each spectrum (see Table \ref{spec_tab}).\\
Estimated errors: wavelength position: $\sim \pm1$\AA~for narrow lines; $\la \pm 10$\AA~for broad lines; FWHM: $\sim \pm 90$ \kms~for the narrow lines, $\la \pm 500$ \kms~ for broad lines; $\phi$: $\sim$ 1-2 \% for narrow lines, $\la 10$\% for broad lines.

\end{flushleft}
\end{table*}

\begin{table*}
\caption{Spectral line parameters as derived from spectra of SN~1996al observed after 100 days.} \label{lines2}
\begin{flushleft}
\begin{tabular}{ccccccc}
\hline
\hline
phase  &          &               &H$\alpha$    &                    & He\,I6678         & He\,I7065\\
(days) &        &         &               &            &           &         \\
\cline{3-5}\\
          &   &B$_{comp}$ &C$_{comp}$ &R$_{comp}$          &        \\

\hline
           & $\lambda_c$ (\AA)                        & 6539             &  6569   & 6597         &  6656   &        \\
+142   & FWHM;V$_{I_0}^{blue}$ (km/s)    &   850;7800    &    950   &   970         &   4600  &        \\
       & $\phi$$^*$                                         &        30          &     26    &    9            &     3      &        \\
\cline{1-2}\\
       & $\lambda_c$ (\AA)                            & 6542            &  6570   & 6595          & 6678   & 7071   \\
+166   & FWHM;V$_{I_0}^{blue}$ (km/s)    &   730;6130   &    680   &   1340       &   2500  &   1800   \\
       & $\phi$$^*$                                         &   30              &     21    &   17           &    4        &  3   \\
\cline{1-2}\\
       & $\lambda_c$ (\AA)                            & 6542             &  6570   & 6594         & 6685    & 7064   \\
+304   & FWHM;V$_{I_0}^{blue}$ (km/s)    &   690;3900     &    800   &   630        &   1950   &   1200   \\
       & $\phi$$^*$                                         &   11                &     11     &    6           &  1.5       &  1.3   \\
\cline{1-2}\\
       & $\lambda_c$ (\AA)                           & 6542             & 6569     & 6593        & 6679     & 7066   \\
+318   & FWHM;V$_{I_0}^{blue}$ (km/s)   &   580;3800    &   880     &   670       &   1100    &   950   \\
       & $\phi$$^*$                                        &    12              &    16      &    7          &  1.0        &  1.5   \\
\cline{1-2}\\
       & $\lambda_c$ (\AA)                           & 6546              & 6573     & 6595        & 6684     & 7070   \\
+406   & FWHM;V$_{I_0}^{blue}$ (km/s)   &   550;3400     &   860     &   680       &   930      &   800   \\
       & $\phi$$^*$                                       &    8                  &    15      &    7          &  3.1        &  4.5   \\
\cline{1-2}\\
       & $\lambda_c$ (\AA)                           & 6544             & 6570     & 6592       & 6680      & 7068   \\
+449   & FWHM;V$_{I_0}^{blue}$ (km/s)   &  510;3200     &   920     &   650       &   850      &   830   \\
       & $\phi$$^*$                                       &    7                 &    16      &    7          &  3.1        &  5.2   \\
\cline{1-2}\\
       & $\lambda_c$ (\AA)                          & 6545              & 6571     & 6592        & 6681      & 7068   \\
+468   & FWHM;V$_{I_0}^{blue}$ (km/s)  &   520,2700     &  1050    &   720        &   950      &   800   \\
       & $\phi$$^*$                                       &    5.5              &    13      &    7           &  2.3        &  3.7   \\
\cline{1-2}\\
       & $\lambda_c$ (\AA)                          & 6546              & 6572      & 6593       & 6682      & 7071   \\
+523   & FWHM;V$_{I_0}^{blue}$ (km/s)  &   520;2400     &   930      &   580       &   980      &   860   \\
       & $\phi$$^*$                                       &    5                 &    9         &    4          &  0.9        &  1.6   \\
\cline{1-2}\\
       & $\lambda_c$ (\AA)                          & 6549             & 6574       & 6597       &  6676     & 7071   \\
+698   & FWHM;V$_{I_0}^{blue}$ (km/s)  &   640;2100    &   860       &   630       &  780       &  860   \\
       & $\phi$$^*$                                       &  2.3               &  2.2         &  1.4         &   0.1:      & 0.2:   \\
\cline{1-2}\\
       & $\lambda_c$ (\AA)                          & 6546             & 6573       & 6598        &               &        \\
+804   & FWHM;V$_{I_0}^{blue}$ (km/s)  &   640;1600    &   1090     &   720        &               &        \\
       & $\phi$$^*$                                       &  1.0               &  1.5         &  0.8          &               &        \\
\cline{1-2}\\
       & $\lambda_c$ (\AA)                                &6549                & 6577             & 6596     &        \\
+1853  & FWHM;V$_{I_0}^{blue}$ (km/s)       &   640;1500    &   820       &   $<630$&  &        \\
       & $\phi$$^*$                                             &  0.9               &  0.6         &  0.2  &        &        \\
\cline{1-2}\\
       & $\lambda_c$ (\AA)                                & 6550    & 6577    & 6594       &        &        \\
+2177  & FWHM;V$_{I_0}^{blue}$ (km/s)      &   960;1700     &   590    &   590       &        &        \\
       & $\phi$$^*$                                            & 0.25     & 0.05      & 0.03        &        &        \\
\cline{1-2}\\
       & $\lambda_c$ (\AA)                                & 6552    & 6577    & 6593       &        &        \\
+2944  & FWHM;V$_{I_0}^{blue}$ (km/s)       &   770;2400    &   680    &   630       &        &        \\
       & $\phi$$^*$                                             & 0.10     & 0.02     & 0.01        &        &        \\
\cline{1-2}\\
       & $\lambda_c$ (\AA)                                & 6551    & 6577    & 6593       &        &        \\
+4178  & FWHM;V$_{I_0}^{blue}$ (km/s)       &   680;2200    &   680    &   630       &        &        \\
       & $\phi$$^*$                                             &  0.13    & 0.02     &0.006       &        &        \\
       
\hline

\end{tabular}

\end{flushleft}
\end{table*}

\begin{table*}
\contcaption{Spectral line parameters as derived from spectra of SN~1996al observed after 100 days.}
\begin{flushleft}
\begin{tabular}{ccccccc}
\hline
\hline
phase    &          &               &H$\alpha$    &                    & He\,I6678         & He\,I7065\\
(days)   &        &         &               &            &           &         \\
\cline{3-5}\\
             & &B$_{comp}$ &C$_{comp}$ &R$_{comp}$    &        &        \\
\hline
       & $\lambda_c$ (\AA)                                & 6552    & 6582    & 6599       &        &        \\
+4850  & FWHM;V$_{I_0}^{blue}$ (km/s)       &   960;1300    &   680    &    180      &        &        \\
       & $\phi$$^*$                                             & 0.13     & 0.03     &0.01         &        &        \\
\cline{1-2}\\
       & $\lambda_c$ (\AA)                                & 6552    & 6575    & 6585       &        &        \\
+5542  & FWHM;V$_{I_0}^{blue}$ (km/s)       &   1000;1500  &   320    &    140      &        &        \\
       & $\phi$$^*$                                             & 0.12     & 0.02     &0.01         &        &        \\

\hline

\end{tabular}

$^*$ In units of $\times 10^{-15}$ erg\,s$^{-1}$\,cm$^{-2}$; not corrected for reddening.\\
\Ha~components better fitted with Lorenztians up to phase +523d; after this phase the \Ha~components are better fitted with Gaussians. He\,I lines always fitted with Gaussians.\\
The FWHMs have been de-convolved for spectral resolution of each spectrum (see Table \ref{spec_tab}). \\
Estimated errors: wavelength position: about $\pm1$\AA~for narrow lines, up to about $\pm 10$\AA~for broad lines; FWHM: about $\pm 90$ \kms~for the narrow lines, up to about $\pm 500$ \kms~ for the broadest lines; $\phi$: about 1-2 \% for narrow lines, up to $\sim 10$\% for the broad lines.

\end{flushleft}
\end{table*}

\end{document}